\documentclass[useAMS,usenatbib]{mn2e}

\usepackage{amsmath}
\usepackage{graphicx}
\usepackage{amssymb}

\newcommand{\lyxdot}{.}

\title[Alignment and signed-intensity anomalies in WMAP data]
{Alignment and signed-intensity anomalies in WMAP data}

\author[Vielva et al.]
{P. Vielva$^{1}$, 
Y. Wiaux$^{2}$, E. Mart{\'\i}nez-Gonz\'alez$^{1}$, P. Vandergheynst$^{2}$\\
$^{1}$Instituto de F{\'\i}sica de Cantabria (CSIC - UC), 39005 Santander, Spain\\
\hspace{0.1cm}E-mails : vielva@ifca.unican.es, martinez@ifca.unican.es \\
$^{2}$Signal Processing Institute, Ecole Polytechnique F\'ed\'erale de Lausanne (EPFL), CH-1015 Lausanne, Switzerland\\
\hspace{0.1cm}E-mails : yves.wiaux@epfl.ch, pierre.vandergheynst@epfl.ch}

\begin{document}

\date{\today}

\pagerange{\pageref{firstpage}--\pageref{lastpage}} \pubyear{2007}

\maketitle

\label{firstpage}

\begin{abstract}
Significant alignment and signed-intensity anomalies of local features
of the cosmic microwave background (CMB) are detected on the three-year
WMAP data, through a decomposition of the signal with steerable wavelets
on the sphere. In addition to identifying local features of a signal
at specific scales, steerable wavelets allow one to determine their
local orientation and signed-intensity. Firstly, an alignment analysis
identifies two mean preferred planes in the sky, both with normal
axes close to the CMB dipole axis. The first plane is defined by the
directions toward which local CMB features are anomalously aligned.
A mean preferred axis is also identified in this plane, located very
close to the ecliptic poles axis. The second plane is defined by the
directions anomalously avoided by local CMB features. This alignment
anomaly provides further insight on recent results \citep{wiaux06a}.
Secondly, a signed-intensity analysis identifies three mean preferred
directions in the southern galactic hemisphere with anomalously high
or low temperature of local CMB features: a cold spot essentially
identified with a known cold spot \citep{vielva04}, a second cold
spot lying very close to the southern end of the CMB dipole axis,
and a hot spot lying close to the southern end of the ecliptic poles
axis. In both analyses, the anomalies are observed at wavelet scales
corresponding to angular sizes around ${10}^{\circ}$ on the celestial
sphere, with global significance levels around $1\%$. Further investigation
reveals that the alignment and signed-intensity anomalies are only
very partially related. Instrumental noise, foreground emissions,
as well as some form of other systematics, are strongly rejected as
possible origins of the detections. An explanation might still be
envisaged in terms of a global violation of the isotropy of the Universe,
inducing an intrinsic statistical anisotropy of the CMB.
\end{abstract}

\begin{keywords}
methods: data analysis, techniques: image processing, cosmology: observations, 
cosmic microwave background
\end{keywords}

\section{Introduction}

\label{sec:introduction} 

In the light of the results obtained from the analysis of several
recent high-precision data sets, a wide consensus was reached in the
cosmology community on a concordance model for a flat $\Lambda\textnormal{CDM}$
Universe with a primordial phase of inflation\emph{.} In this framework,
the Universe is flat and filled in with cold dark matter (CDM) and
dark energy in the form of a cosmological constant ($\Lambda$), in
addition to the standard baryonic and electromagnetic components.
The cosmological principle assumption states that the Universe is
globally homogeneous and isotropic. The large scale structure and
the cosmic microwave background (CMB) radiation find their origin
in Gaussian quantum energy density fluctuations developed around the
homogeneous and isotropic background, as predicted by the standard
inflationary scenario taking place in the primordial Universe.

The CMB data, and specifically the data provided by the NASA Wilkinson
Microwave Anisotropy Probe (WMAP) satellite experiment, have played
a leading role in defining this concordance model \citep{bennett03a,spergel03,hinshaw07,spergel07}.
However anomalies have been reported. They naturally question the
basic hypotheses on which the model relies, in particular the Gaussianity
(assumed by the standard inflationary scenario) and statistical isotropy
(postulated by the cosmological principle) of the statistical distribution
from which the CMB temperature fluctuations arise in the primordial
Universe.

Departures from Gaussianity of the one-year and three-year WMAP data
have been reported in terms of various statistics. Firstly, non-Gaussianity
was notably detected through the use of a genus-based statistic \citep{park04}.
Secondly, wavelet analyses have also reported non-Gaussian deviations.
An excess in the kurtosis of the wavelet coefficients of the axisymmetric
Mexican hat wavelet on the sphere was found at wavelet scales corresponding
to angular sizes on the celestial sphere around $10^{\circ}$ \citep{vielva04}.
This deviation is located in the southern galactic hemisphere, where
a very cold spot was identified at $(\theta,\varphi)=(147^{\circ},209^{\circ})$,
with $\theta\in[0,\pi]$ and $\varphi\in[0,2\pi)$ respectively standing
for the co-latitude and longitude in galactic spherical coordinates.
This detection was confirmed with various wavelets and various statistics
\citep{mukherjee04,cruz05,cayon05,mcewen05,cruz06,cruz07,mcewen06}.
Notice that the cold spot identified not only bears non-Gaussianity,
but also represents a departure from statistical isotropy, in terms
of a North-South asymmetry in galactic coordinates.

Departures from statistical isotropy of the one-year and three-year
WMAP data have also been reported in terms of various statistics.

Firstly, a North-South asymmetry in ecliptic coordinates has been
detected through the use of N-point correlation functions \citep{eriksen04a,eriksen05},
Minkowski functionals \citep{eriksen04b}, local power spectra \citep{hansen04b,donoghue05},
local bispectra \citep{land05a}, and local curvature \citep{hansen04a}.
The asymmetry is maximum in a coordinate system with the North pole
lying at $(\theta,\varphi)=(80^{\circ},57^{\circ})$, close to the
northern end of the ecliptic poles axis at $(\theta,\varphi)=(60^{\circ},96^{\circ})$.
This North-South asymmetry was confirmed through the application of
a pair angular-separation histogram method \citep{bernui06,bernui07}.
More recently, it was again observed through the search for a best-fit
dipolar modulation of the three-year WMAP data \citep{spergel07,eriksen07}.
The corresponding dipolar axis has a northern end at $(\theta,\varphi)=(63^{\circ},45^{\circ})$.

Secondly, an anomalous alignment of the lowest multipoles of the WMAP
data was reported, especially at $\ell=2,3$ \citep{deOliveira04,schwarz04,copi04,katz04,bielewicz05},
but also up to $\ell=5$ \citep{land05b,abramo06}, and for $\ell=6,7$
\citep{freeman06}. This alignment highlights the so-called axis of
evil, with northern end at $(\theta,\varphi)=(30^{\circ},260^{\circ})$,
very close to the CMB dipole axis, with northern end at $(\theta,\varphi)=(42^{\circ},264^{\circ})$.
Very recently, a new statistic to analyze the alignment of multipoles
that is robust against treatments of the galactic plane was proposed
\citep{land06}. The corresponding analysis highlights a weaker detection
for these alignments.

The large scale fluctuations of the CMB were also pointed out to fit
with the expected pattern of an anisotropic Bianchi VII$_{h}$ Universe
\citep{jaffe06a,jaffe06c,bridges06}, although the cosmological parameters
derived from this hypothesis are ruled-out \citep{jaffe06b,bridges06}.
Notice on the contrary that bipolar power spectra analyses are consistent
with no violation of the statistical isotropy of the Universe \citep{hajian05b,hajian03,hajian05a}.

Finally, in a previous work, we detected a very significant deviation
from statistical isotropy of the one-year WMAP data through the use
of a novel alignment analysis, based on the decomposition of the WMAP
data with steerable wavelets on the sphere \citep{wiaux06a}. This
alignment analysis probes the anomalous alignment of local CMB features
toward specific directions on the celestial sphere. At wavelet scales
corresponding to angular sizes around ${10}^{\circ}$, an anomalous
alignment is observed toward a mean preferred plane in the sky, whose
normal axis, with a northern end at $(\theta,\varphi)=(34^{\circ},331^{\circ})$,
is close to the CMB dipole axis and close to the axis of evil. In
addition, a prominent cluster of directions toward which local features
are aligned identifies a mean preferred axis in this plane, with a
northern end at $(\theta,\varphi)=(71^{\circ},91^{\circ})$, very
close to the ecliptic poles axis. The detection of this alignment
anomaly actually synthesized the previously reported statistical anisotropy
results by highlighting both the ecliptic poles axis and the CMB dipole
axis as preferred axes in the sky. But further analyses were still
required to establish their origin.

The aim of this paper is to probe further the statistical isotropy
of the CMB through two different analyses. Firstly, we reproduce,
on the three-year WMAP data, the previously discussed alignment analysis
of the local CMB features originally performed on the one-year WMAP
data. Secondly, we propose a signed-intensity analysis of the local
CMB features, based on the same steerable wavelet decomposition of
the three-year WMAP data. This \emph{}signed-intensity analysis probes
the anomalously high or low temperature of local CMB features. The
paper is organized as follows. In Section \ref{sec:methodology},
we discuss the overall methodologies defined for the subsequent analyses.
In Section \ref{sec:analyses}, we present the results of analysis
of the three-year WMAP data. In Section \ref{sec:questfororigin},
we investigate possible origins for the anomalies observed. We finally
conclude in Section \ref{sec:conclusion}.

\section{Methodologies}

\label{sec:methodology}

In this section, we firstly recall the notion of wavelet steerability
and present the particular steerable wavelet on the sphere used in
the subsequent analysis. Secondly, we discuss how to differentiate
between statistical anisotropy and non-Gaussianity detections through
the analysis of the CMB data with simulations produced following the
concordance model. We finally recall the procedure for the alignment
analysis, and define the new procedure for the signed-intensity analysis,
both intended to probe the statistical isotropy of the CMB data.

\subsection{Steerable wavelets}

A practical approach to the formalism of continuous wavelets on the
sphere (\emph{i.e.} the unit sphere $S^{2}$), and the corresponding
notion of steerable wavelets were recently introduced \citep{wiaux05}.
Let us primarily fix some notations. We consider a three-dimensional
Cartesian coordinate system $(o,o\hat{x},o\hat{y},o\hat{z})$ centered
on the sphere, and where the direction $o\hat{z}$ identifies the
North pole. Any point $\omega$ on the sphere is identified by its
corresponding spherical coordinates $(\theta,\varphi)$, where $\theta\in[0,\pi]$
stands for the co-latitude, and $\varphi\in[0,2\pi)$ for the longitude.
Any filter invariant under rotation around itself is said to be axisymmetric.
By definition, any non-axisymmetric, or directional, filter is steerable
if a rotation by $\chi\in[0,2\pi)$ around itself may be expressed
in terms of a finite linear combination of non-rotated basis filters.
In this work, we consider the second Gaussian derivative wavelet (2GD),
$\Psi^{\partial_{\hat{x}}^{2}(gau)}$, which is obtained by a stereographic
projection of the second derivative in direction $\hat{x}$ of a Gaussian
in the tangent plane at the North pole. The filter obtained is a steerable
wavelet on the sphere which may be rotated in terms of three basis
filters: the second derivative in the direction $\hat{x}$ itself,
$\Psi^{\partial_{\hat{x}}^{2}(gau)}$ , the second derivative in the
direction $\hat{y}$, $\Psi^{\partial_{\hat{y}}^{2}(gau)}$, and the
cross-derivative, $\Psi^{\partial_{\hat{x}}\partial_{\hat{y}}(gau)}$
\citep{wiaux05}. These basis filters dilated at any scale $a\in\mathbb{R}_{+}^{*}$
read: 

\begin{eqnarray}
\Psi_{a}^{\partial_{\hat{x}}^{2}(gau)}\left(\theta,\varphi\right) & = & f_{a}\left(\theta\right)\left[1-\frac{4}{a^{2}}\tan^{2}\frac{\theta}{2}\cos^{2}\varphi\right]\label{eq:2gdbasis1}\\
\Psi_{a}^{\partial_{\hat{y}}^{2}(gau)}\left(\theta,\varphi\right) & = & f_{a}\left(\theta\right)\left[1-\frac{4}{a^{2}}\tan^{2}\frac{\theta}{2}\sin^{2}\varphi\right]\label{eq:2gdbasis2}\\
\Psi_{a}^{\partial_{\hat{x}}\partial_{\hat{y}}(gau)}\left(\theta,\varphi\right) & = & f_{a}\left(\theta\right)\left[-\frac{2}{a^{2}}\tan^{2}\frac{\theta}{2}\sin2\varphi\right],\label{eq:2gdbasis3}\end{eqnarray}
 where $f_{a}(\theta)=a^{-1}\sqrt{4/3\pi}(1+\tan^{2}(\theta/2))e^{-2\tan^{2}(\theta/2)/a^{2}}$.
The value of the scale $a$ identifies with the dispersion of the
Gaussian in units of $2\tan(\theta/2)$. The angular size of the 2GD
is defined as twice the half-width of the wavelet, where the half-width
is defined by $\theta_{hw}=2\arctan(a/2)$, which is closely approximated
by $a$ at small scales.

The analysis of a signal $F$ with a given wavelet $\Psi$ simply
defines a set of wavelet coefficients $W_{\Psi}^{F}(\omega_{0},\chi,a)$,
which result from the directional correlation between $F$ and the
wavelet dilated at any scale $a$, $\Psi_{a}$. In other words these
wavelet coefficients are defined by the scalar product between the
signal and the wavelet dilated at scale $a$, rotated around itself
by $\chi$, and translated at any point $\omega_{0}$ on the sphere,
also denoted $\Psi_{\omega_{0},\chi,a}$: \begin{equation}
W_{\Psi}^{F}(\omega_{0},\chi,a)=\langle\Psi_{\omega_{0},\chi,a}|F\rangle=\int_{S^{2}}d\Omega\Psi_{\omega_{0},\chi,a}^{*}(\omega)F(\omega).\label{eq:direccorr}\end{equation}
 The $^{*}$ denotes complex conjugation. The wavelet coefficients
of a signal therefore characterize the signal at each scale $a$,
orientation $\chi$, and position $\omega_{0}$. By linearity of the
operation defining wavelet coefficients from the wavelet, the relation
of steerability of the wavelet is automatically transferred to the
wavelet coefficients. The specific steerability relation for the wavelet
coefficients of a signal $F$ with the 2GD reads:\begin{eqnarray}
W_{\Psi^{\partial_{\hat{x}}^{2}}}^{F}(\omega_{0},\chi,a) & = & W_{\Psi^{\partial_{\hat{x}}^{2}}}^{F}\left(\omega_{0},a\right)\cos^{2}\chi\nonumber \\
 &  & +W_{\Psi^{\partial_{\hat{y}}^{2}}}^{F}\left(\omega_{0},a\right)\sin^{2}\chi\nonumber \\
 &  & +W_{\Psi^{\partial_{\hat{x}}\partial_{\hat{y}}}}^{F}\left(\omega_{0},a\right)\sin2\chi,\label{eq:steerability}\end{eqnarray}
where $W_{\Psi^{\partial_{\hat{x}}^{2}}}^{F}(\omega_{0},a)$, $W_{\Psi^{\partial_{\hat{y}}^{2}}}^{F}(\omega_{0},a)$,
and $W_{\Psi^{\partial_{\hat{x}}\partial_{\hat{y}}}}^{F}(\omega_{0},a)$
are the wavelet coefficients respectively obtained from the standard
correlation (\emph{i.e.} $\chi=0$) of the signal with $\Psi_{a}^{\partial_{\hat{x}}^{2}(gau)}$,
$\Psi_{a}^{\partial_{\hat{y}}^{2}(gau)}$ and $\Psi_{a}^{\partial_{\hat{x}}\partial_{\hat{y}}(gau)}$.

At each scale $a$ and at each position $\omega_{0}$, one is typically
interested in one or a few of the infinite number of wavelet coefficients
associated with the infinite number of orientations $\chi$ \citep{wiaux05,mcewen07}.
Selecting the orientation $\chi_{0}(\omega_{0},a)$ that maximizes
the absolute value of the wavelet coefficient, corresponds to selecting
the local orientation at which the wavelet best matches the local
feature of the signal. Notice that, as the 2GD is invariant under
rotation around itself by $\pi$, orientations may arbitrarily be
constrained in a range of length $\pi$. As $\Psi_{a}^{\partial_{\hat{x}}^{2}(gau)}$
oscillates in the tangent direction $\hat{x}$, it actually detects
features aligned along the tangent direction $\hat{y}$. Hence, the
local orientation of the feature itself, $D^{F}(\omega_{0},a)$, is
defined in terms of $\chi_{0}=\chi_{0}(\omega_{0},a)$ as: \begin{equation}
\frac{\pi}{2}\leq D^{F}(\omega_{0},a)=\chi_{0}+\frac{\pi}{2}<\frac{3\pi}{2}.\label{eq:orientation}\end{equation}
For the particular case of the 2GD, the steerability relation (\ref{eq:steerability})
enables to compute the orientation $\chi_{0}$ from the wavelet coefficients
obtained from the basis filters as: \begin{equation}
\tan2\chi_{0}=\frac{2W_{\Psi^{\partial_{\hat{x}}\partial_{\hat{y}}}}^{F}\left(\omega_{0},a\right)}{W_{\Psi^{\partial_{\hat{x}}^{2}}}^{F}\left(\omega_{0},a\right)-W_{\Psi^{\partial_{\hat{y}}^{2}}}^{F}\left(\omega_{0},a\right)}.\label{eq:2gdchi0}\end{equation}
 The wavelet coefficient itself at scale $a$, position $\omega_{0}$,
and in direction $\chi_{0}$, defines to so-called signed-intensity
of the local feature:\begin{equation}
I^{F}(\omega_{0},a)=W_{\Psi^{\partial_{\hat{x}}^{2}}}^{F}(\omega_{0},\chi_{0},a).\label{eq:signed-intensity}\end{equation}

Analyzing signals with steerable wavelets is interesting in several
respects. Firstly, the wavelet decomposition enables one to identify
the scales $a$ of the physical processes which define the local feature
of the signal at each point $\omega_{0}$. Secondly, the steerability
theoretically gives access to the orientation and signed-intensity
of these local features. Finally, from the computational point of
view, the calculation of a directional correlation at each analysis
scale is an extremely demanding task. The relation of steerability
is essential to reduce the complexity of calculation of the wavelet
coefficients when local orientations are considered \citep{wiaux06b}.

\subsection{Statistical anisotropy or non-Gaussianity?}

As already discussed in the introduction, under the cosmological principle
assumption and in the framework of the standard inflationary scenario,
the CMB signal can be interpreted as a realization of a Gaussian and
statistically isotropic random field.

It is not a trivial task to develop an analysis procedure for testing
statistical properties with only one realization (because there is
only one Universe) of the CMB signal. Typically, the data are contaminated
by noise and foreground emissions. Simulations accounting for these
peculiarities must be performed to test the compatibility of the data
with a given set of assumptions. There are infinite ways of simulating
departures from Gaussianity or statistical isotropy. But unless one
is interested in the compatibility of the data with a specific alternative
model (considering for example a non-standard inflationary scenario,
an anisotropic Universe, a non-trivial topology of the Universe, topological
defects, etc.), simulations are produced following the concordance
model. In that context, any statistical incompatibility of the data
with the simulations is to be interpreted as a departure from the
whole assumption of Gaussianity and statistical isotropy. Despite
that, postulating Gaussianity allows one to interpret a detection
as a departure from statistical isotropy, and conversely. Typically,
analyses using local estimators on the sphere, for example local power
spectra, can naturally identify preferred directions and probe statistical
isotropy, provided that Gaussianity is postulated. On the contrary,
analyses using global estimators on the sphere, for example statistical
moments computed by averages on the whole sphere, explicitly assume
statistical isotropy in order to probe Gaussianity.

In this paper, two analysis procedures are considered. The alignment
analysis is recalled, which probes the anomalous alignment of local
CMB features toward specific directions. A \emph{}new signed-intensity
analysis is also defined, which probes the anomalously high or low
temperature of local CMB features. Because of the local nature of
the corresponding anomalies on the celestial sphere, we use these
analyses as specific probes of the statistical isotropy of the CMB,
while its Gaussianity is postulated.

\subsection{Alignment analysis procedure}

\label{sub:Alignment-analysis-procedure}

The alignment of local CMB features toward specific directions on
the celestial sphere can be probed with steerable wavelets by combining
at each analysis scale $a$ and at each point $\omega_{0}$ the information
on the orientation $D^{T}(\omega_{0},a)$ and on the signed-intensity
$I^{T}(\omega_{0},a)$ (the superscript $^{T}$ identifies the CMB
temperature field) \citep{wiaux06a,vielva06}. Firstly, the great circle
is defined which passes by the point $\omega_{0}$ and admits as a
tangent the local direction defined by $D^{T}(\omega_{0},a)$. All
directions on that great circle are considered to be seen by the local
feature at $\omega_{0}$ with a weight naturally given as the absolute
value of the signed-intensity $\vert I^{T}(\omega_{0},a)\vert$ at
that point. At scale $a$ and in an arbitrary direction $\omega$,
the total weight $TW^{T}(\omega,a)$ is defined as the sum of the
$N_{cros}(\omega)$ weights originating from all pixels $\omega_{0}^{(c)}$
in the original CMB signal, with $1\leq c\leq N_{cros}(\omega)$,
for which the great circle defined crosses the direction considered:
\begin{equation}
TW^{T}\left(\omega,a\right)=\frac{1}{A}\sum_{c=1}^{N_{cros}(\omega)}\vert I^{T}\left(\omega_{0}^{(c)},a\right)\vert.\label{eq:total-weights}\end{equation}
 The factor $A=LN_{pix}^{-1}\sum_{\omega_{0}\notin M_{a}}\vert I^{T}(\omega_{0},a)\vert$
simply stands for the normalization of the total weights. It takes
into account the total number of pixels on the sphere $N_{pix}$ and
the mean number $L$ of pixels on each great circle in the pixelization
considered. It also accounts for the presence of an exclusion mask
$M_{a}$ containing the pixels to be excluded from the analysis at
each scale $a$. As discussed in the next section, our analyses are
performed on HEALPix pixelizations%
\footnote{http://healpix.jpl.nasa.gov/%
} \citep{gorski05}. In this scheme, at a given resolution $N_{side}$,
the number of pixels on each great circle is constant and reads $L=4N_{side}-1$.
These total weights constitute a new signal on the celestial sphere.
Notice that the procedure obviously assigns identical total weights
to opposite directions, as any great circle on the sphere always contains
a direction and its opposite direction on the same axis. In other
words, the total weights signal is even under parity in the three-dimensional

\begin{figure}
\includegraphics[width=8cm,keepaspectratio]{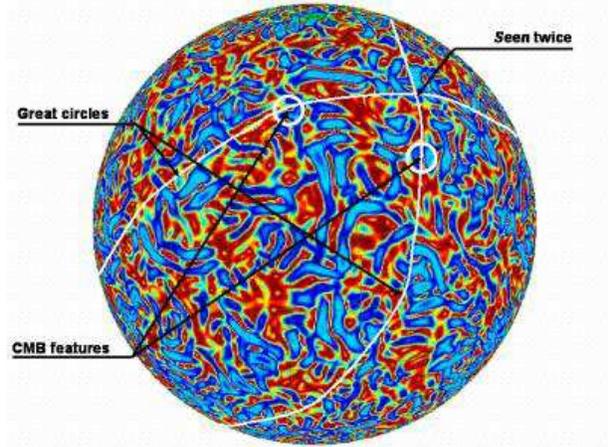}

\caption{\label{fig:scheme}Alignment analysis procedure: example of signed-intensity
distribution on the celestial sphere. At each point $\omega_{0}$,
the great circle is defined which passes by $\omega_{0}$ and admits
as a tangent the direction of the local feature at that point, defined
by $D^{T}(\omega_{0},a)$. All directions on that great circle are
considered to be seen by the local CMB feature at $\omega_{0}$, with
a weight naturally given as the absolute value of the signed-intensity
$\vert I^{T}(\omega_{0},a)\vert$ at that point. At each analysis
scale $a$, the total weight $TW^{T}(\omega,a)$ in each direction
$\omega$ is defined as the sum of the weights originating from all
pixels for which the great circle defined crosses the direction considered.
The total weights signal on the celestial sphere is even under parity.
In an arbitrary hemisphere of reference, the biggest total weights
identify the directions toward which the CMB features are predominantly
aligned, while the lowest total weights identify the directions predominantly
avoided by the CMB features.}
\end{figure}

Cartesian coordinate system $(o,o\hat{x},o\hat{y},o\hat{z})$ centered
on the sphere. The analysis of this signal can be restricted to the
directions $\omega$ in one arbitrary hemisphere of reference. In
that hemisphere, the biggest total weights identify the directions
toward which the CMB features are predominantly aligned, while the
lowest total weights identify the directions predominantly avoided
by the CMB features. An illustration of the procedure described is
given in Figure \ref{fig:scheme}. %

The signal of total weights is analyzed by comparison with the values
obtained from simulations%
\footnote{Notice that analyzed maps are intrinsically anisotropic. This is due
to the statistical anisotropy of the noise and to the presence of
a mask that cuts the brighest foreground emissions (see Subsection
\ref{sub:datasimul}). The statistical distribution of local quantities
must therefore be computed independently at each pixel, through the
means of simulations.%
} at the same scale $a$ and in the same direction $\omega$. Ten thousand
simulations were produced for the analysis. At each point $\omega$
of the pixelization considered, the cumulative probability that the
simulations exhibit a lower value than the simulation with the $i^{th}$
lowest value is $(i-0.5)\times10^{-2}\%$. The cumulative probability
that the simulations exhibit a lower value than the data is computed
by linear interpolation between the cumulative probabilities assigned
to the two neighbour simulations with values directly surrounding
the value of the data. A value of the data lower than all simulation
values is assigned a cumulative probability of $0.005\%$, while a
value of the data higher than all simulation values is assigned a
cumulative probability of $99.995\%$. Let $\mu_{TW}^{T}(\omega,a)$
be the median total weight value predicted by the Gaussian and statistically
isotropic simulations. Directions with a total weight higher than
the median value of the simulations, $TW^{T}(\omega,a)\geq\mu_{TW}^{T}(\omega,a)$,
are said to bear a positive total weight. Directions with a total
weight lower than the median value of the simulations, $TW^{T}(\omega,a)\leq\mu_{TW}^{T}(\omega,a)$,
are said to bear a negative total weight. For positive total weights,
a direction $\omega$ is defined to be anomalous at a percentage $p$
if the corresponding cumulative probability that the simulations exhibit
a lower value than the data is higher or equal to $p$. For negative
total weights, a direction $\omega$ is defined to be anomalous at
a percentage $p$ if the corresponding cumulative probability that
the simulations exhibit a lower value than the data is lower or equal
to $1-p$. In the following, the precise threshold $p=99.865\%$ is
considered because it formally corresponds to the $+3\sigma$ and
$-3\sigma$ values in a Gaussian distribution, for positive and negative
total weights respectively. In other words, a direction $\omega$
with a positive total weight value of the data higher or equal to
the $9987^{th}$ lowest simulation value (\emph{i.e.} at maximum $14$
simulations exhibit a value higher or equal) is considered to be anomalous
at $99.865\%$. A direction $\omega$ with a negative total weight
value of the data lower or equal to the $14^{th}$ lowest simulation
value (\emph{i.e.} at maximum $14$ simulations exhibit a value lower
or equal) is also considered to be anomalous at $99.865\%$.

The global significance level for an alignment detection is estimated
as the percentage of simulations with a number of anomalous directions
higher or equal to the number of anomalous directions in the data.
The lower the significance level, the stronger the detection. Notice
that instrumental noise or spurious emissions might produce isolated
anomalous directions, corresponding to isolated anomalous pixels.
The estimation of the global significance level corrects for that
effect in the total count, by weighting isolated anomalous pixels
with a low clustering index, and weighting anomalous pixels surrounded
by other anomalous pixels, \emph{i.e.} clustered anomalous directions,
with a high clustering index. This clustering index is calculated
as the fraction of anomalous pixels in a disk with a diameter equal
to the angular size of the wavelet at scale $a$.

\subsection{Signed-intensity analysis procedure}

\label{sub:Signed-intensity-analysis-procedure}

Alternatively, the statistical isotropy of the CMB signal can be probed
with steerable wavelets by simply considering at each analysis scale
$a$ the signed-intensity $I^{T}(\omega_{0},a)$ of the local CMB
features at each point $\omega_{0}$ on the celestial sphere.

The signed-intensity signal (see Figure \ref{fig:scheme}) is also
analyzed by comparison with the values obtained from simulations at
the same scale $a$ and in the same direction $\omega$. At each point
$\omega$ of the pixelization considered, the cumulative probability
that the simulations exhibit a lower value than the data is computed
through the exact same procedure as for the alignment analysis, by
linear interpolation from the values of the ten thousand simulations.
The median signed-intensity value predicted by the Gaussian and statistically
isotropic simulations is zero in each direction as the CMB temperature
fluctuations are defined with zero statistical mean. Positive signed-intensities,
$I^{T}(\omega_{0},a)\geq0$, therefore indicate directions with a
temperature higher than the median value of the simulations. Negative
signed-intensities, $I^{T}(\omega_{0},a)\leq0$, indicate directions
with a temperature lower than the median value of the simulations.
Again, for positive signed-intensities, a direction $\omega$ is defined
to be anomalous at a percentage $p$ if the corresponding cumulative
probability that the simulations exhibit a lower value than the data
is higher or equal to $p$. For negative signed-intensities, a direction
$\omega$ is defined to be anomalous at a percentage $p$ if the corresponding
cumulative probability that the simulations exhibit a lower value
than the data is lower or equal to $1-p$. We still take $p=99.865\%$,
corresponding to the $+3\sigma$ and $-3\sigma$ values in a Gaussian
distribution, for positive and negative signed-intensities respectively.

The global significance level for a signed-intensity detection is
also estimated as the percentage of simulations with a number of anomalous
directions higher or equal to the number of anomalous directions in
the data, still weighting each anomalous pixel by its clustering index.

\section{WMAP analyses}

\label{sec:analyses} 

In this section, we firstly discuss the WMAP three-year data and simulations
used for our analyses. Secondly, we present the results of the alignment
and signed-intensity analyses.

\subsection{Data and simulations}

\label{sub:datasimul}

Firstly, we consider the data. We have analyzed the three-year WMAP
data after correction for foreground emissions contamination by a
template fitting technique \citep{spergel07}. This procedure provides
eight cleaned maps at various frequencies for the corresponding WMAP
radiometers: Q1 and Q2 at $41$ GHz, V1 and V2 at $61$ GHz, and W1,
W2, W3, and W4 at $94$ GHz. These maps are available from the NASA
LAMBDA archive%
\footnote{http://lambda.gsfc.nasa.gov/%
}. After foreground emissions removal, these maps are masked with the
Kp0 mask \citep{spergel07} that cuts the regions ($\approx20\%$ of
the sky) with the brightest galaxy emission and the positions ($\approx5\%$
of the sky) where the brightest known point sources are located, assigning
zero values to the corresponding pixels. It is assumed that the CMB
is convolved with an instrumental beam that is well described by an
isotropic window function and that the instrumental noise is to first
order Gaussian, statistically anisotropic, and uncorrelated. Maps
with better signal-to-noise ratio can be obtained, by an optimal combination
of the eight cleaned maps obtained after foreground emissions removal
and masking. At each pixel, this combination is obtained by weighting
each map by the corresponding inverse noise variance to produce the
so-called three-year WMAP co-added CMB map \citep{hinshaw07}, denoted
here the WCM123 map. For the sake of the analysis, this map is considered
to contain only CMB and instrumental noise. The residual foreground
emissions are considered to be negligible.

In addition to the WCM123 map, other maps are produced for different
analysis purposes. Three maps are built to optimize signal-to-noise
ratio at each frequency by combining the foreground cleaned and masked
maps of the corresponding radiometers: the WCM-Q, WCM-V, and WCM-W.
These frequency maps obviously contain CMB and instrumental noise,
as well as residual foreground emissions which are precisely not neglected
by principle. They are actually used to test whether detected anomalies
are due to residual foreground emissions, which are frequency-dependent.
Three other maps are also produced to remove in depth both CMB and
residual foreground emissions at each frequency by subtracting the
foreground cleaned and masked maps of the corresponding radiometers:
Q1 - Q2 defines the WCM-nQ map, V1 - V2 defines the WCM-nV maps, and
W1 - W2 + W3 - W4 defines the WCM-nW map. Despite negligible CMB and
foreground emissions residuals at small angular sizes (well below
$1^{\circ}$) due to the different beam window functions of the radiometers
at each frequency, these difference maps essentially contain instrumental
noise. They are used to test whether detected anomalies are due to
any effect associated with instrumental noise. A last map is produced
to remove in depth only CMB by subtracting the foreground cleaned
and masked maps at different frequencies, defining the WCM-nWVQ map
as W1 + W2 + W3 + W4 -V1 - V2 - Q1 - Q2. Despite negligible CMB residuals
at small angular sizes (well below $1^{\circ}$) again due to the
different beam window functions of the radiometers at each frequency,
this difference map essentially contains instrumental noise and residual
foreground emissions. It can be used as an additional test of the
impact of instrumental noise and residual foreground emissions on
detected anomalies.

In order to minimize any contamination coming from errors on the cosmological
dipole subtraction, the dipole outside the mask is removed from each
analyzed map \citep{komatsu03}.

All these maps are initially produced in HEALPix pixelization at the
resolution $N_{side}=512$, corresponding to maps with several million
equal-area pixels with a spatial resolution of $6.9'$. For the sake
of our analysis, which is applied at wavelet scales corresponding
to angular sizes between $5^{\circ}$ and $30^{\circ}$, they are
downgraded to the resolution $N_{side}=32$. This provides maps with
$N_{pix}=12288$ equal-area pixels with a spatial resolution of $1.8^{\circ}$.

Secondly, we consider the simulations. Each of the simulations used
to compare the results of the analysis of the data to what is expected
from the concordance model was produced as follows. The spherical
harmonics coefficients of a Gaussian and statistically isotropic CMB
realization are obtained from the angular power spectrum determined
by the cosmological parameters of the three-year WMAP best-fit model
\citep{spergel07} with CAMB (Code for Anistropies in the Microwave
Background%
\footnote{http://camb.info/%
}). The observation at each of the eight WMAP radiometers considered
is simulated by convolving that signal in harmonic space with the
corresponding isotropic window function. Each map is then transformed
to pixel space at the resolution $N_{side}=512$, and a Gaussian and
statistically anisotropic noise realization is added with the proper
dispersion per pixel. This provides simulations of the CMB signal,
as seen by the radiometers at the different WMAP frequencies considered.
The same prescriptions as those described above for the data are then
applied to produce one simulated coadded map (WCM123), three simulated
frequency maps (WCM-Q, WCM-V, and WCM-W), and three simulated difference
maps (WCM-nQ, WCM-nV, WCM-nW, and WCM-nWVQ). These simulations are
then also downgraded at the resolution $N_{side}=32$ for our analysis.

Ten thousand simulations of the WCM123 map were produced for the main
analysis. In order to reduce the corresponding computation time, only
one thousand simulations were considered for the frequency maps and
difference maps used to analyze the origin of a possible detection
in terms of instrumental noise or foreground emissions. This is by
far enough since no precise significance level is established from
the analysis of these maps. Only the general patterns of possible
detections are studied in comparison with the patterns observed from
the analysis of the WCM123 map.

Finally, in valid pixels close to masked regions, the result of the
directional correlation of a signal with steerable wavelets at a given
scale of analysis is inevitably affected by the zero values of the
Kp0 mask. On each of the maps presented above, and obviously both
for the data and simulations, an exclusion mask $M_{a}$ is therefore
defined at each wavelet scale $a$ in order to exclude the affected
pixels from the analysis \citep{vielva04}.

\subsection{Alignment analysis}

\begin{figure}
\includegraphics[width=8cm,keepaspectratio]{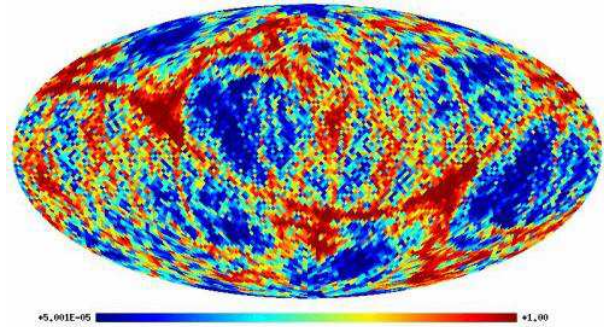}

\caption{\label{fig:TW}Alignment analysis: cumulative probabilities map in
Mollweide projection for the total weights $TW^{T}(\omega,a_{3})$
computed from the WCM123 map, at wavelet scale $a_{3}$ corresponding
to ${8.3}^{\circ}$ of angular size. The pattern observed presents
several great circles of high positive (in red) or negative (in blue)
total weights on the celestial sphere, respectively corresponding
to directions with a total weight value well higher, or lower, than
the median value of the simulations.}
\end{figure}

We have performed the alignment analysis on the WCM123 map with the
2GD at twelve wavelet scales $a_{i}$ ($1\leq i\leq12$) corresponding
to angular sizes between $5^{\circ}$ and $30^{\circ}$. The wavelet
half-widths $\theta_{hw}$ corresponding to the scales $a_{1}$ to
$a_{12}$ respectively read in arcminutes as: \{$150'$, $200'$,
$250'$, $300'$, $350'$, $400'$, $450'$, $500'$, $550'$, $600'$,
$750'$, $900'$\}. The map of the cumulative probabilities that the
simulations exhibit a lower value than the data for the total weights
$TW^{T}(\omega,a_{3})$ at wavelet scale $a_{3}$, corresponding to
$8.3^{\circ}$ of angular size, is presented in Figure \ref{fig:TW}.
One can easily acknowledge that the pattern of direction preferences
presents several great circles on the celestial sphere. At that scale,
an anomaly in the alignment of the local CMB features is actually
observed, which is detailed in the following.%

\begin{figure}
\includegraphics[height=8cm,keepaspectratio,angle=-90]{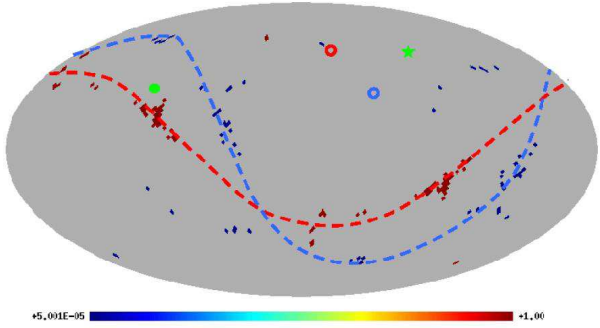}

\caption{\label{fig:perR03tw}Alignment analysis: thresholded cumulative probabilities
map in Mollweide projection for the total weights $TW^{T}(\omega,a_{3})$
computed from the WCM123 map, at wavelet scale $a_{3}$ corresponding
to ${8.3}^{\circ}$ of angular size. The total weights signal is even
under parity, and its analysis can be restricted to one hemisphere
of reference. Considering for instance the northern galactic hemisphere,
$67$ directions anomalous at $99.865\%$ are retained. The global
significance level of detection is $0.85\%$. The directions with
positive total weights (in red) define a first mean preferred plane
(red dashed line) in the sky, with a normal axis (red circle) close
to the CMB dipole axis (green asterisk). In this plane, a prominent
cluster of directions identifies a mean preferred axis very close
to the ecliptic poles axis (green disk). The directions with negative
total weights (in blue) define a second mean preferred plane (blue
dashed line) with a normal axis (blue circle) again close to the CMB
dipole axis.}
\end{figure}

Let us recall that the total weights signal is even under parity,
and its analysis can be restricted to one hemisphere of reference.
Considering for instance the northern galactic hemisphere, $67$ directions
anomalous at $99.865\%$ are identified at wavelet scale $a_{3}$.
These anomalous directions are represented in Figure \ref{fig:perR03tw}.
The global significance level of detection is $0.85\%$. Among these
directions, $39$ exhibit positive total weights, corresponding to
directions with a total weight value well higher than the median value
of the simulations. They are essentially aligned along a great circle,
which defines a first mean preferred plane in the sky, toward which
local CMB features are anomalously aligned. The normal axis to this
plane, whose northern end is identified by $(\theta,\varphi)=(39^{\circ},337^{\circ})$
in galactic co-latitude $\theta$ and longitude $\varphi$, lies close
to the CMB dipole axis, with northern end at $(\theta,\varphi)=(42^{\circ},264^{\circ})$,
and close to the axis of evil, with northern end at $(\theta,\varphi)=(30^{\circ},260^{\circ})$.
In this plane, a prominent cluster of anomalous directions toward
which local features are anomalously aligned identifies a mean preferred
axis whose northern end lies at $(\theta,\varphi)=(73^{\circ},91^{\circ})$,
very close to the ecliptic poles axis, with northern end at $(\theta,\varphi)=(60^{\circ},96^{\circ})$.
The $28$ remaining anomalies exhibit negative total weights, corresponding
to directions with a total weight value well lower than the median
value of the simulations. They are also essentially aligned along
a great circle, which defines a second mean preferred plane in the
sky, of directions anomalously avoided by the local CMB features.
The normal axis to this plane, whose northern end is identified by
$(\theta,\varphi)=(59^{\circ},309^{\circ})$, lies also close to the
CMB dipole axis and to the axis of evil. The whole structure of the
present detection hence provides further insight into the anomaly
recently observed \citep{wiaux06a} on the one-year WMAP co-added CMB
map \citep{bennett03b}. It again synthesizes the previously reported
statistical anisotropy results by highlighting both the ecliptic poles
axis and the CMB dipole axis as preferred axes in the sky. %

The detection is confirmed at wavelet scale $a_{4}$, corresponding
to $10^{\circ}$ of angular size. At this scale, in one hemisphere
of reference, $62$ directions anomalous at $99.865\%$ are identified.
The global significance level of detection is $1.28\%$. Among these
directions, $38$ exhibit positive total weights, and the $24$ remaining
exhibit negative total weights. They define the same pattern of anomaly
as the one observed at wavelet scale $a_{3}$.

\subsection{Signed-intensity analysis}

\begin{figure}
\includegraphics[width=8cm,keepaspectratio]{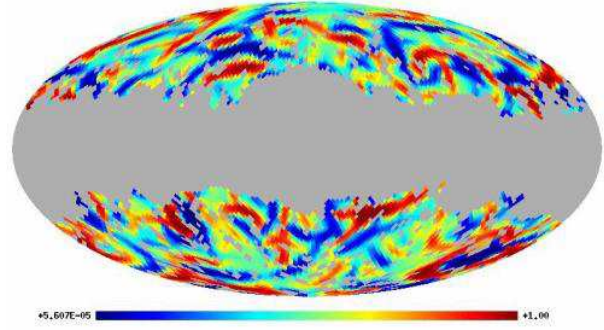}

\caption{\label{fig:WC}Signed-intensity analysis: cumulative probabilities
map in Mollweide projection for the signed-intensities $I^{T}(\omega,a_{3})$
computed for the WCM123 map, at wavelet scale $a_{3}$ corresponding
to ${8.3}^{\circ}$ of angular size. The pattern observed presents
several well located spots of high positive (in red) or negative (in
blue) signed-intensities on the celestial sphere, respectively corresponding
to directions with a temperature value well higher, or lower, than
the median value of the simulations.}
\end{figure}

We have performed the signed-intensity analysis on the WCM123 map
with the 2GD at the same scales as for the alignment analysis, corresponding
to angular sizes between $5^{\circ}$ and $30^{\circ}$. The map of
the cumulative probabilities that the simulations exhibit a lower
value than the data for the signed-intensities $I^{T}(\omega_{0},a_{3})$
at wavelet scale $a_{3}$ corresponding to ${8.3}^{\circ}$ of angular
size, is presented in Figure \ref{fig:WC}. The pattern observed is
actually similar to the one obtained through the axisymmetric Mexican
hat wavelet analysis \cite[Figure 12]{vielva04}, presenting several
well located spots of high positive or negative signed-intensities
on the celestial sphere. Beyond that similarity, at that scale, an
anomaly in the signed-intensity of the local CMB features is observed,
which is detailed in the following.%

\begin{figure}
\includegraphics[height=8cm,keepaspectratio,angle=-90]{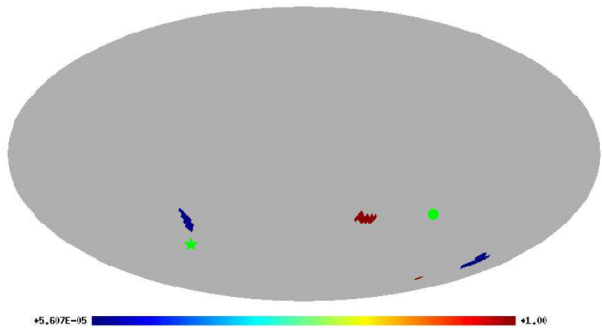}

\caption{\label{fig:perR03wc}Signed-intensity analysis: thresholded cumulative
probabilities map in Mollweide projection for the signed-intensities
$I^{T}(\omega,a_{3})$ computed for the WCM123 map, at wavelet scale
$a_{3}$ corresponding to ${8.3}^{\circ}$ of angular size. On the
whole celestial sphere, $39$ directions anomalous at $99.865\%$
are retained. The global significance level of detection is $1.39\%$.
The anomalous directions are distributed in three clusters in the
southern galactic hemisphere, identifying three mean preferred directions
in the sky. A first cold spot (in blue) identifies with a known cold
spot \citep{vielva04}. A second cold spot (in blue) lies very close
to the southern end of the CMB dipole axis (green asterisk). The third
spot is a hot spot (in red) close to the southern end of the ecliptic
poles axis (green disk). }
\end{figure}

Considering here the whole celestial sphere and of course not only
one hemisphere, $39$ directions anomalous at $99.865\%$ are identified
at wavelet scale $a_{3}$. These anomalous directions are represented
in Figure \ref{fig:perR03wc}. The global significance level of detection
is $1.39\%$. Among these directions, $24$ exhibit negative signed-intensities,
corresponding to directions with a temperature value well lower than
the median value of the simulations. The $15$ remaining anomalies
exhibit positive signed-intensities, corresponding to directions with
a temperature value well higher than the median value of the simulations.
The anomalous directions are actually distributed in three clusters
in the southern galactic hemisphere, identifying three mean preferred
directions in the sky. A first cold spot (\emph{i.e.} of negative
signed-intensities) centered at $(\theta,\varphi)=(150^{\circ},209^{\circ})$
in galactic co-latitude $\theta$ and longitude $\varphi$, essentially
identifies with the known cold spot centered at $(\theta,\varphi)=(147^{\circ},209^{\circ})$
\citep{vielva04}. A second cold spot is centered at $(\theta,\varphi)=(129^{\circ},80^{\circ})$,
very close to the southern end of the CMB dipole axis at $(\theta,\varphi)=(138^{\circ},84^{\circ})$.
The third spot is a hot spot (\emph{i.e.} of positive signed-intensities)
centered at $(\theta,\varphi)=(124^{\circ},321^{\circ})$, close to
the southern end of the ecliptic poles axis, at $(\theta,\varphi)=(120^{\circ},276^{\circ})$.
In conclusion, this detection again synthesizes previously reported
statistical anisotropy results. Indeed, it confirms the North-South
asymmetry, to which the cold spot centered at $(\theta,\varphi)=(147^{\circ},209^{\circ})$
was originally related. Moreover, two other spots are identified in
the southern galactic hemisphere which contribute to the North-South
asymmetry, and with specific positions once more close to both the
ecliptic poles axis and the CMB dipole axis. %

The detection is confirmed at wavelet scales $a_{4}$, $a_{5}$, and
$a_{6}$, respectively corresponding to $10^{\circ}$, $11.7^{\circ}$,
and $13.3^{\circ}$ of angular sizes. The three spots are explicitly
recovered, with global significance levels below $4.90\%$. In particular
at wavelet scale $a_{4}$, $42$ directions anomalous at $99.865\%$
are identified on the whole celestial sphere, among which $29$ exhibit
negative signed-intensities, and $13$ exhibit positive signed-intensities.
The global significance level of detection is $1.97\%$.

Let us also emphasize the peculiar distribution on the celestial sphere
of the three spots detected, which are roughly separated two by two
by right angles. They identify the three basis vectors of a coordinate
frame with the North pole close to the southern end of the ecliptic
poles axis. The hot spot coincides with the North pole, and the cold
spots roughly lie in the perpendicular plane, with one of them very
close to the southern end of the CMB dipole axis.

\subsection{All-scale significance level}

An all-scale significance level may be computed, which should account
for the fact that twelve wavelet scales have been probed, hence artificially
increasing chances of detection at one or another scale. But it should
also account for all the detections observed in both the alignment
and the signed-intensity analyses. In the data, the two alignment
detections at the wavelet scales $a_{3}$ and $a_{4}$, and the four
signed-intensity detections at the wavelet scales $a_{3}$, $a_{4}$,
$a_{5}$, and $a_{6}$ are observed at global significance levels
roughly lower than $5.00\%$. In a conservative approach, the all-scale
significance level defined is estimated as the percentage of simulations
with a global significance level lower or equal to $5.00\%$ for at
least two consecutive scales in the alignment analysis, and at least
four consecutive scales (not necessarily related with the first two)
in the signed-intensity analysis. The corresponding value for this
robust all-scale significance level is $1.50\%$. This actually confirms
the best levels of detections observed in the data in terms of global
significance levels at individual wavelet scales in each of the two
analyses.

\section{anomalies origin}

\label{sec:questfororigin}

In this section, we firstly establish that the alignment and signed-intensity
anomalies detected are only very partially related. Secondly, we consider
and discard possible origins of these anomalies in terms of instrumental
noise, foreground emissions, and some form of unknown systematics.

\subsection{Anomalies relation}

The alignment and signed-intensity anomalies both highlight the ecliptic
poles axis and the CMB dipole axis as preferred axes in the sky. Moreover,
in terms of procedure, the alignment analysis based on the definition
(\ref{eq:total-weights}) of total weights can be interpreted as a
variation of the signed-intensity analysis, taking into account the
orientation of local CMB features. Consequently, it is natural to
raise the question of a possible relation between the corresponding
anomalies detected. More precisely, one can verify if the part of
the alignment anomaly associated with positive total weights originates,
at least partially, in the signed-intensity anomaly. In order to probe
this hypothesis, the alignment analysis was repeated on the WCM123
map, in which each of the three spots responsible for the signed-intensity
anomaly was in turn excluded from the analysis\emph{.} In that regard,
the pixels corresponding to the spot considered were simply included
in the exclusion mask.

When the hot spot close to the ecliptic poles axis is excluded, the
number of directions anomalous at $99.865\%$ associated with positive
total weights drops, in one hemisphere of reference, from $39$ to
$28$ at wavelet scale $a_{3}$, and the corresponding global significance
level increases from $0.87\%$ to $2.38\%$. At wavelet scale $a_{4}$,
the number of directions anomalous at $99.865\%$ associated with
positive total weights drops, again in one hemisphere of reference,
from $38$ to $18$, and the corresponding global significance level
increases more drastically from $1.66\%$ to $17.31\%$. On the contrary,
there is no significant impact on the alignment anomaly of neither
of the two cold spots responsible for the signed-intensity anomaly.
In other words, the hot spot of anomalous signed-intensities is partially,
but certainly not totally, responsible for the part of the alignment
anomaly associated with positive total weights, while the two cold
spots have no significant impact on it. The observed alignment anomaly
associated with positive total weights (see Figure \ref{fig:perR03tw})
is therefore not due to the alignment of a small number of local CMB
features with an anomalous signed-intensity. It is rather due to the
alignment of a large number of local CMB features with a non-anomalous
signed-intensity. This result is in agreement with a previous analysis
showing that the local CMB features aligned toward the directions
exhibiting anomalous positive total weights are essentially homogeneously
distributed on the celestial sphere \citep{vielva06}. In conclusion,
the alignment anomaly observed originates only very partially in the
signed-intensity anomaly.

\subsection{Instrumental noise and foreground emissions}

Before claiming a possible cosmological origin of the detections reported,
it is essential to look for possible origins in terms of instrumental
noise or foreground emissions.

\begin{figure*}
\includegraphics[width=4cm,keepaspectratio]{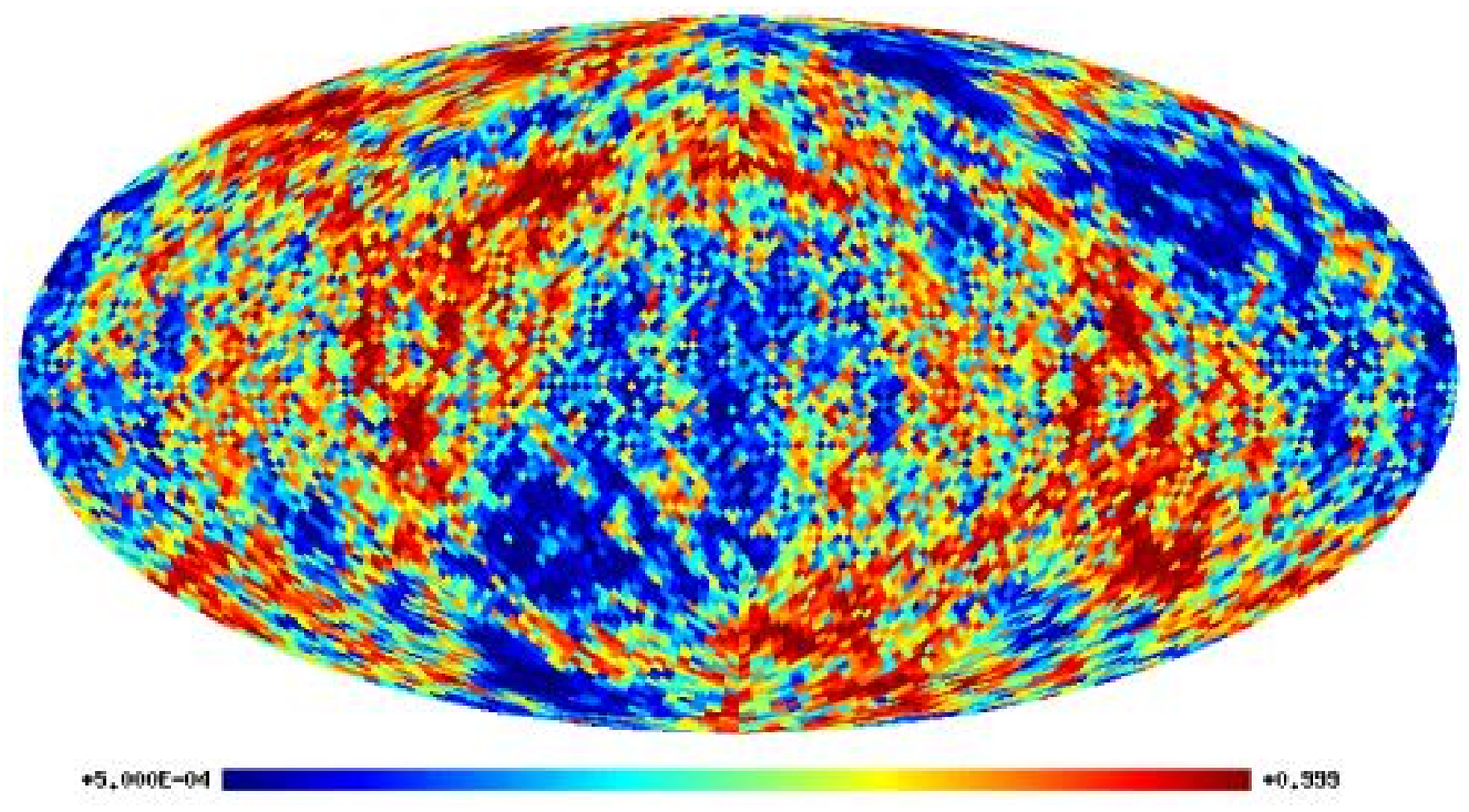}\includegraphics[width=4cm,keepaspectratio]{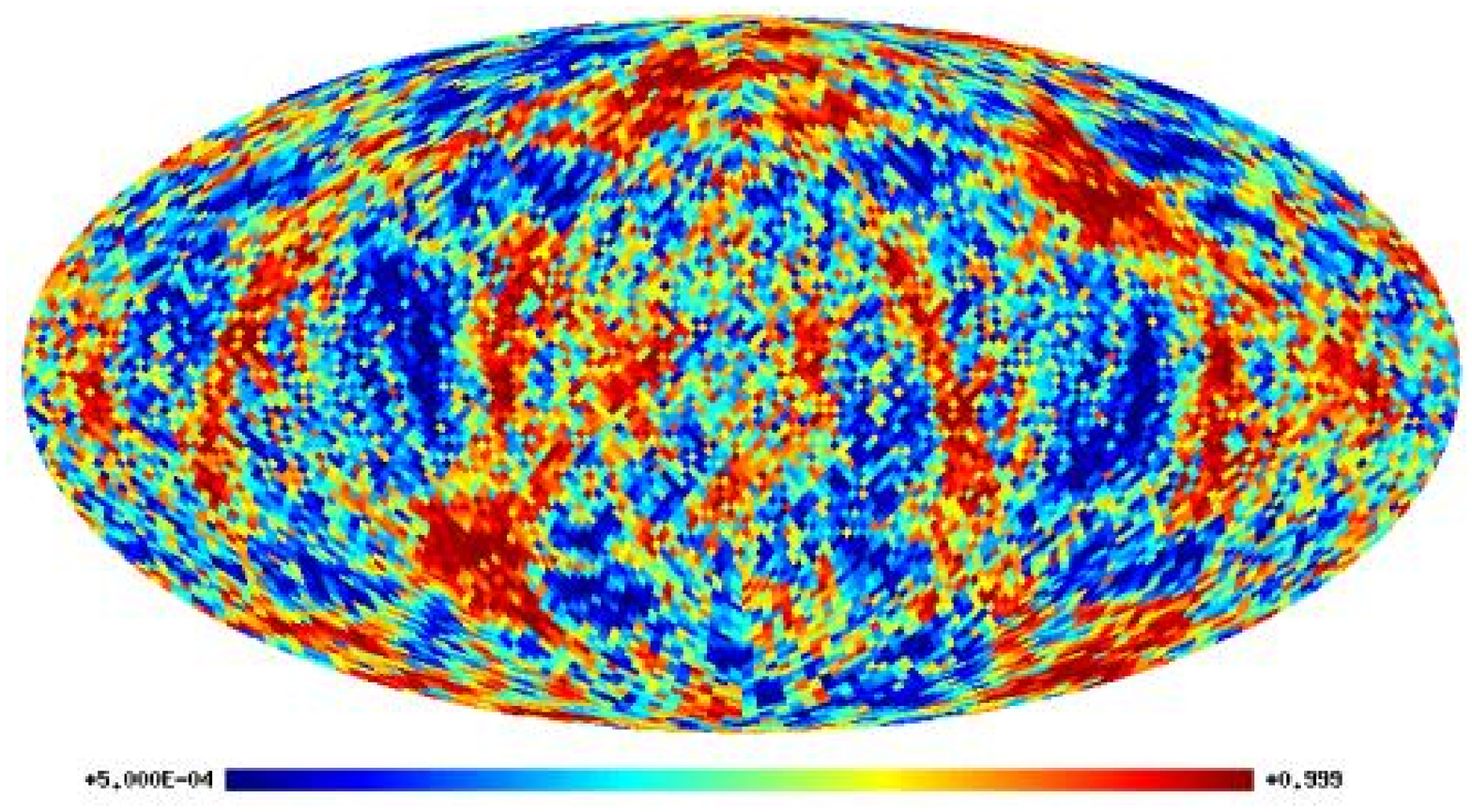}\includegraphics[width=4cm,keepaspectratio]{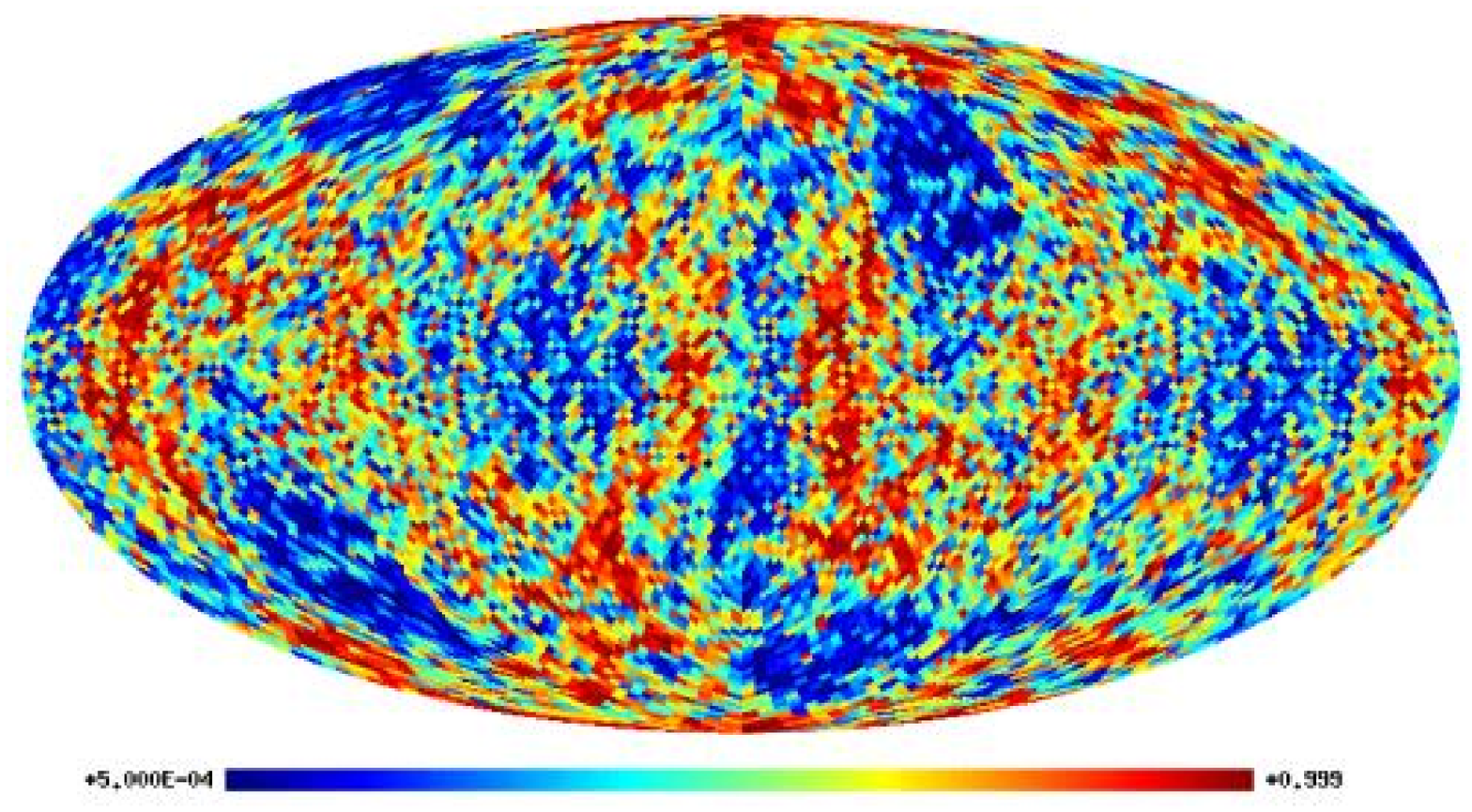}

\includegraphics[width=4cm,keepaspectratio]{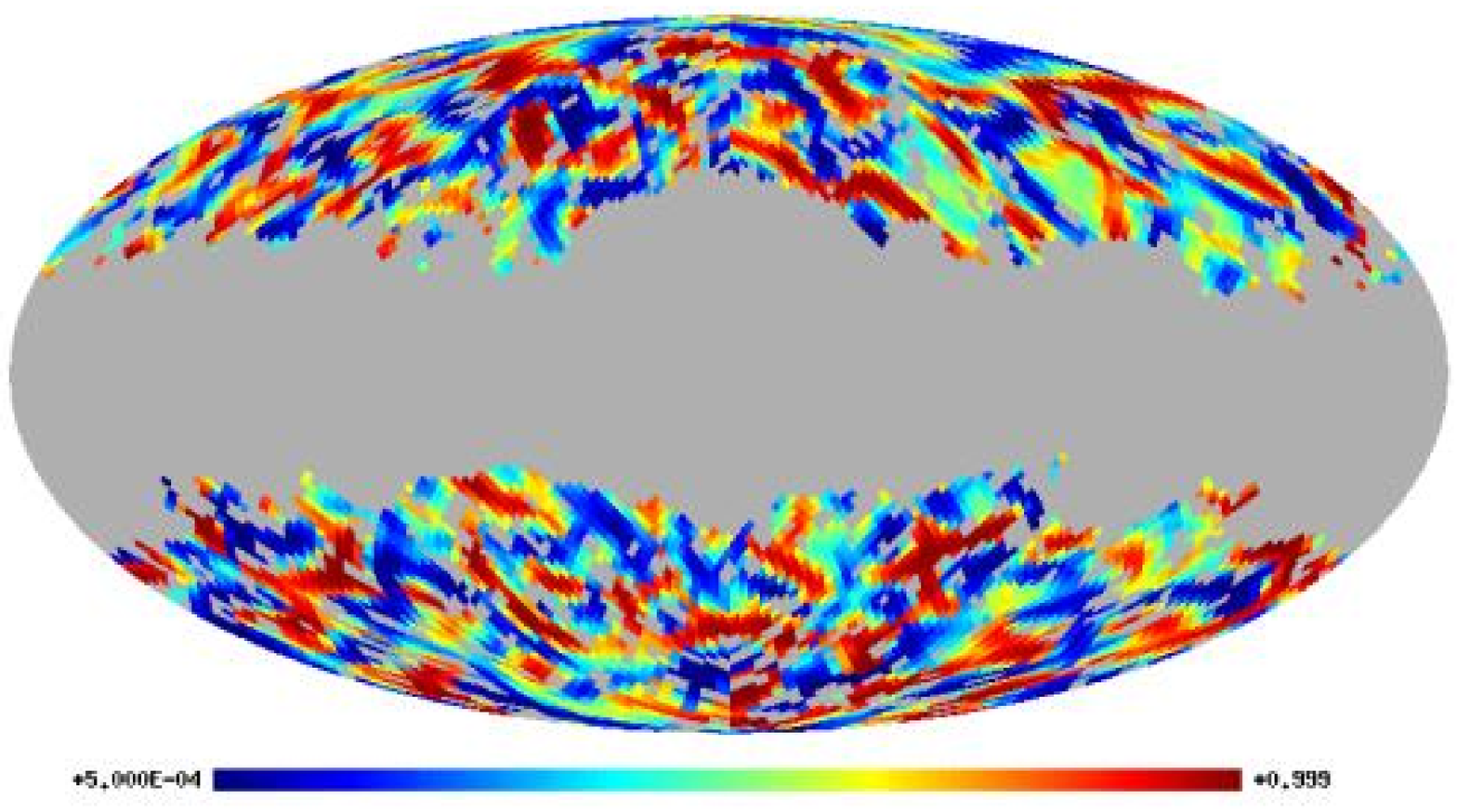}\includegraphics[width=4cm,keepaspectratio]{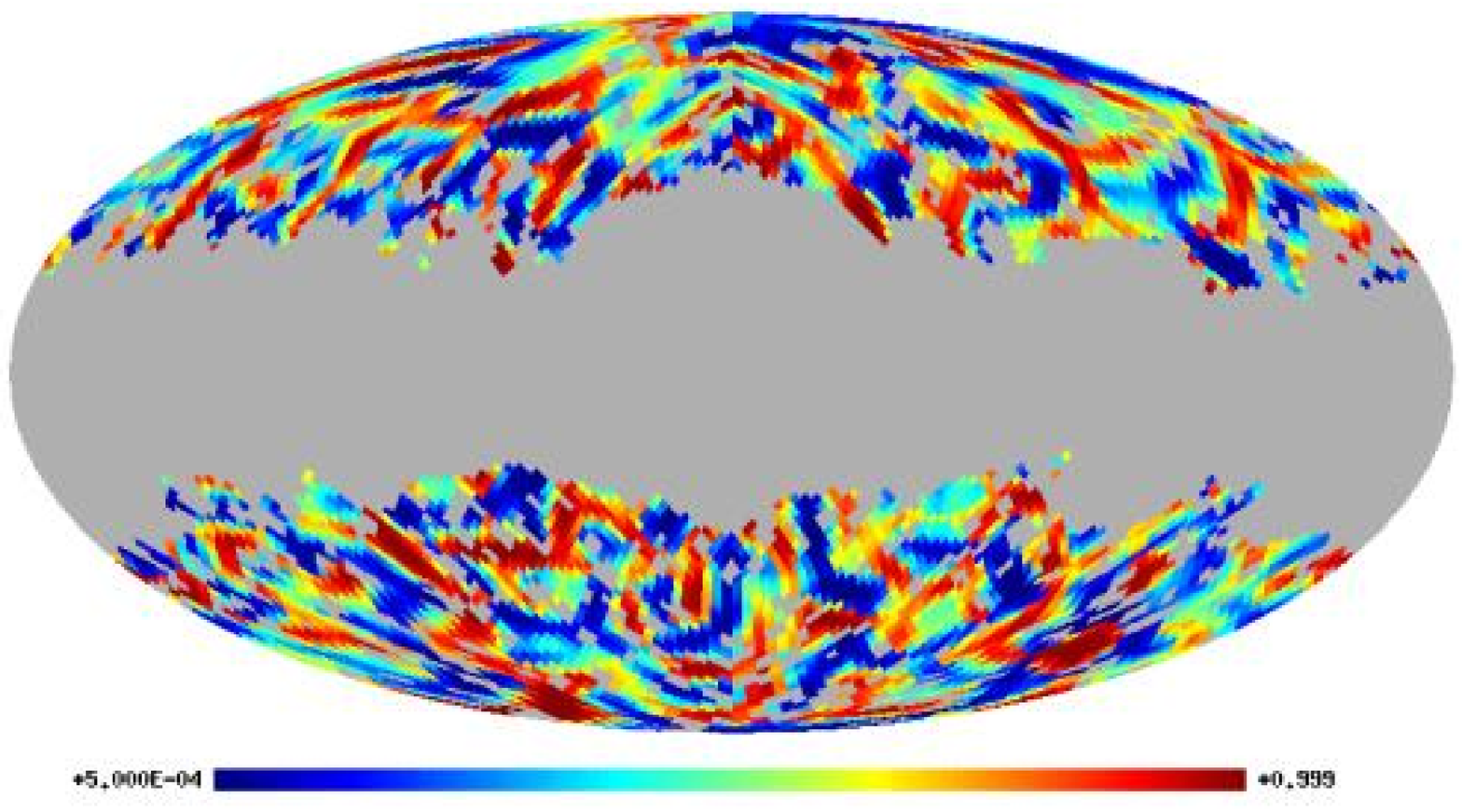}\includegraphics[width=4cm,keepaspectratio]{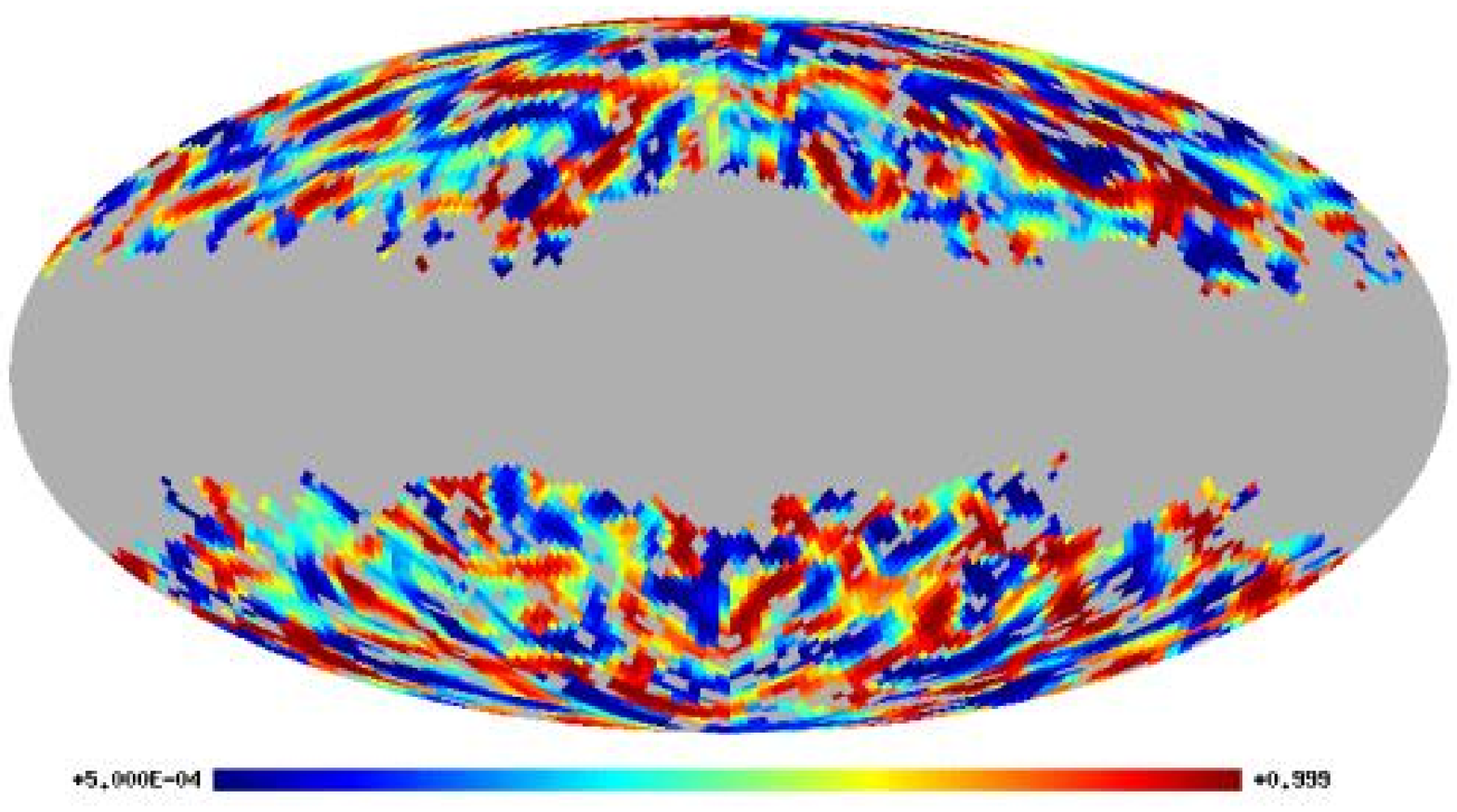}

\caption{\label{fig:noise}Alignment and signed-intensity analyses: cumulative
probabilities maps in Mollweide projection for the total weights $TW^{T}(\omega,a_{3})$
(top panel) and signed-intensities $I^{T}(\omega,a_{3})$ (bottom
panel) at wavelet scale $a_{3}$, resulting from the difference maps
WCM-nQ (left column), WCM-nV (center column), and WCM-nW (right column).
The total weights and signed-intensities patterns do not correspond
to those obtained from the WCM123 map. Instrumental noise can be therefore
be discarded as a possible origin of the alignment and signed-intensity
detections.}
\end{figure*}

A possible origin in terms of instrumental noise in the WCM123 map
is analyzed by reproducing the analyses on the difference maps WCM-nQ,
WCM-nV, and WCM-nW, which are essentially noise maps (see Subsection
\ref{sub:datasimul}). In particular, it is known that the WMAP data
present a small component of correlated noise at large angular sizes
(well above $1^{\circ}$), which is not taken into account in our
simulations. This instrumental noise might be detected through the
analysis of these difference maps. Moreover, the pattern of this correlated
noise is related to the ecliptic coordinates system through the WMAP
scanning strategy. It is therefore a plausible origin of the alignment
and signed-intensity anomalies observed. The corresponding maps representing
the cumulative probabilities that the simulations exhibit a lower
value than the data are presented in Figure \ref{fig:noise}, both
for the total weights $TW^{T}(\omega,a_{3})$ and signed-intensities
$I^{T}(\omega,a_{3})$ at wavelet scale $a_{3}$. The total weights
and signed-intensities patterns revealed do not correspond to those
obtained from the WCM123 map on which the anomalies are observed.
Instrumental noise can therefore be discarded as a possible origin
of the alignment and signed-intensity detections.%

\begin{figure*}
\includegraphics[width=4cm,keepaspectratio]{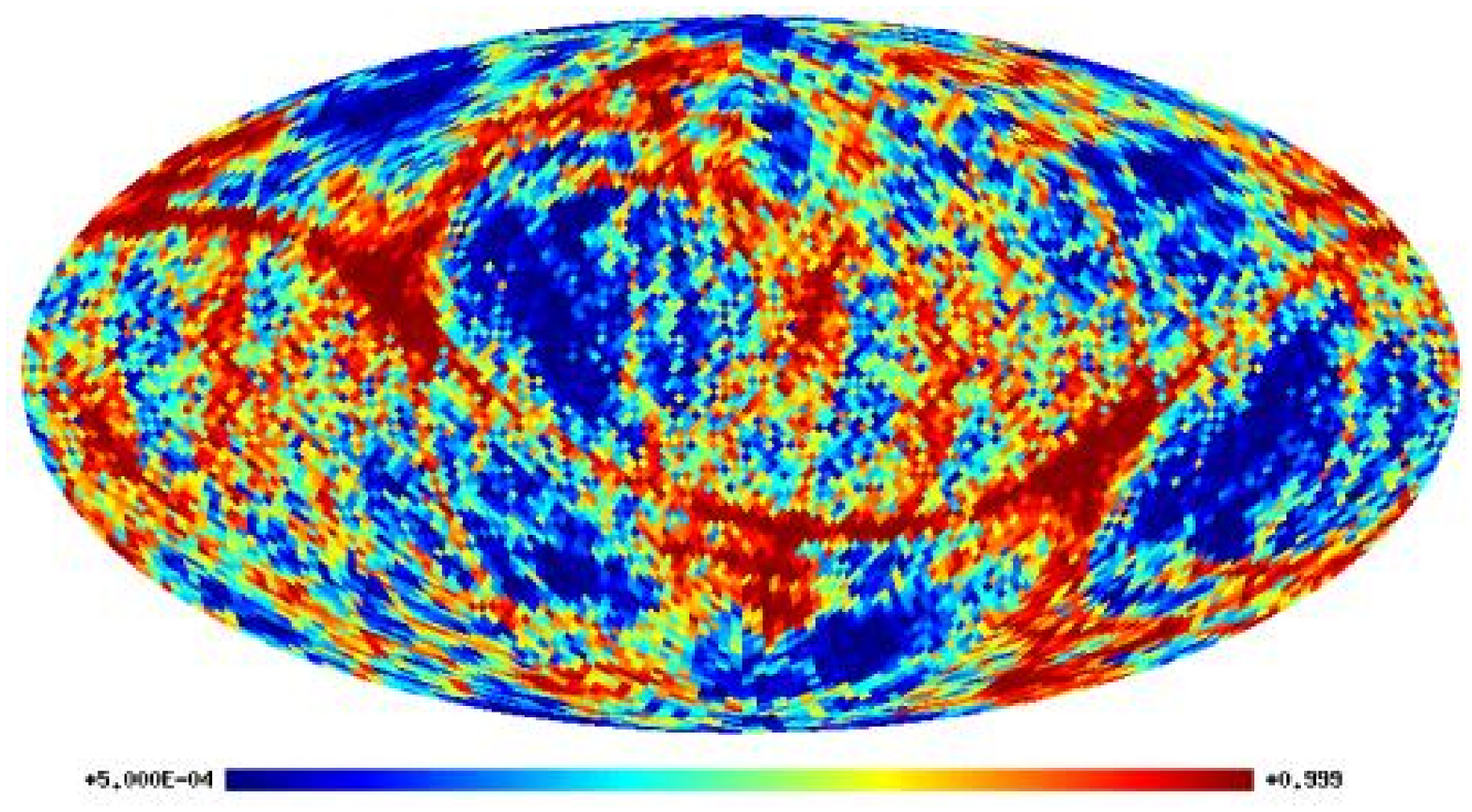}\includegraphics[width=4cm,keepaspectratio]{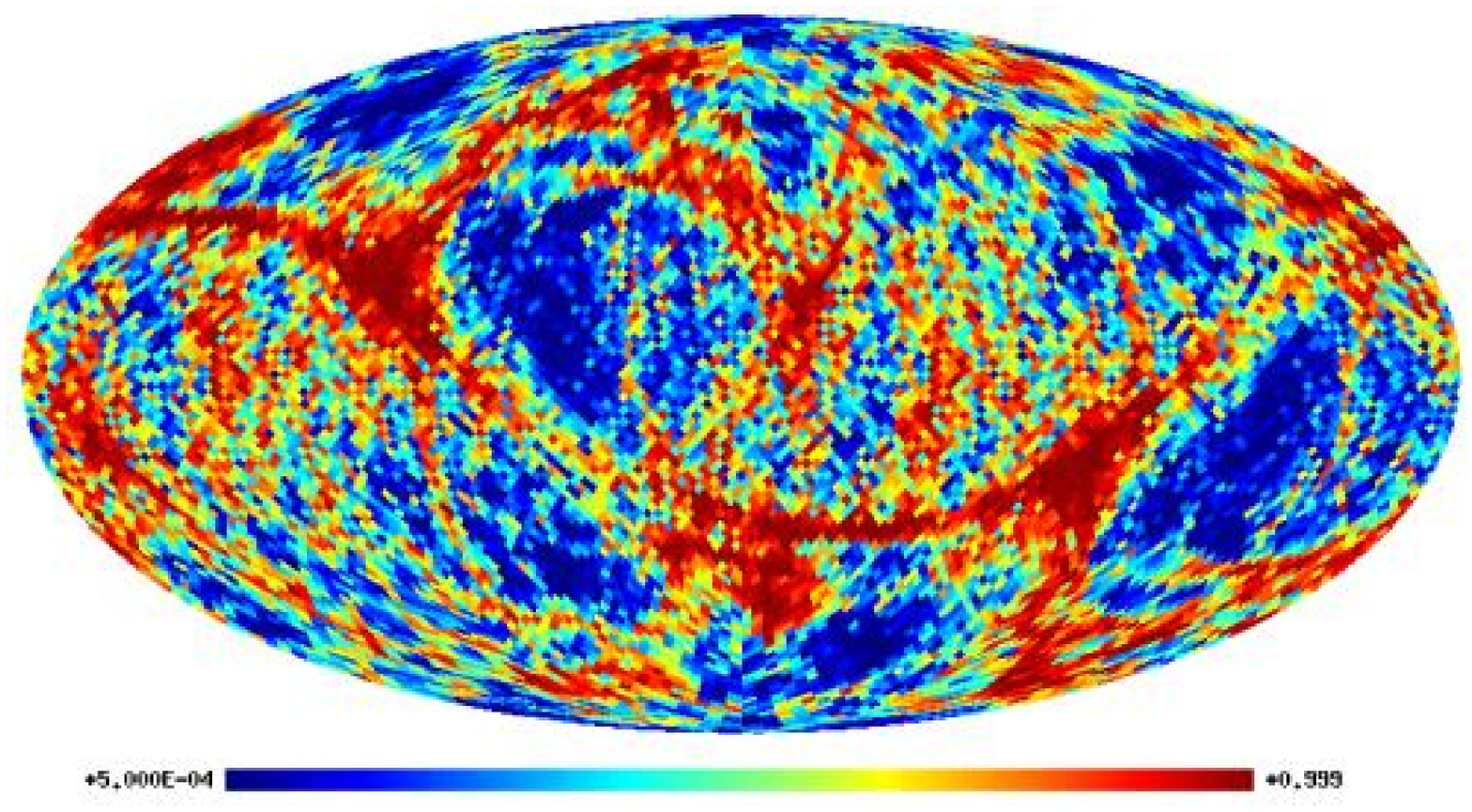}\includegraphics[width=4cm,keepaspectratio]{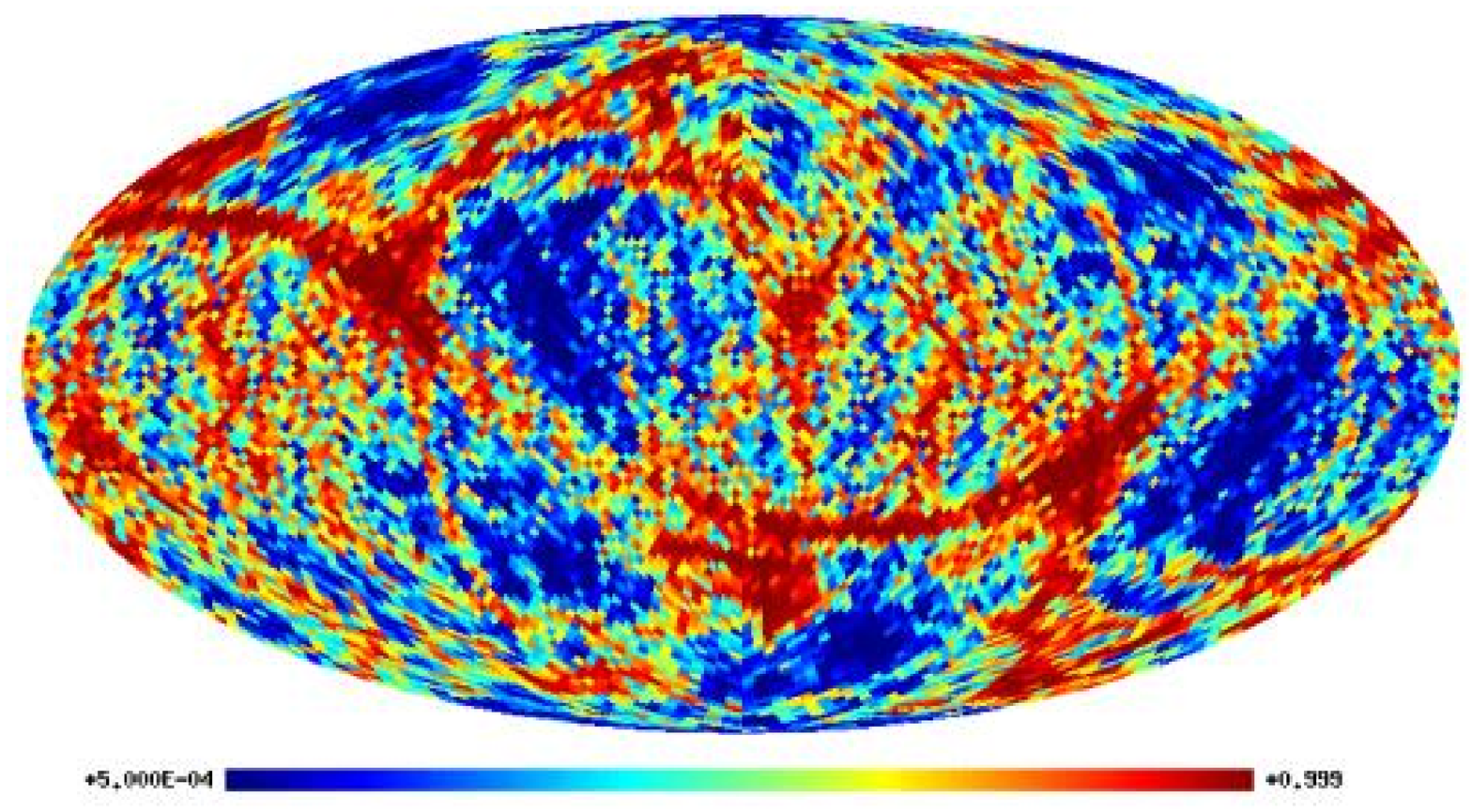}\includegraphics[width=4cm,keepaspectratio]{./percentage_pixels_R03_Totalweights-W1W2W3W4}

\includegraphics[width=4cm,keepaspectratio]{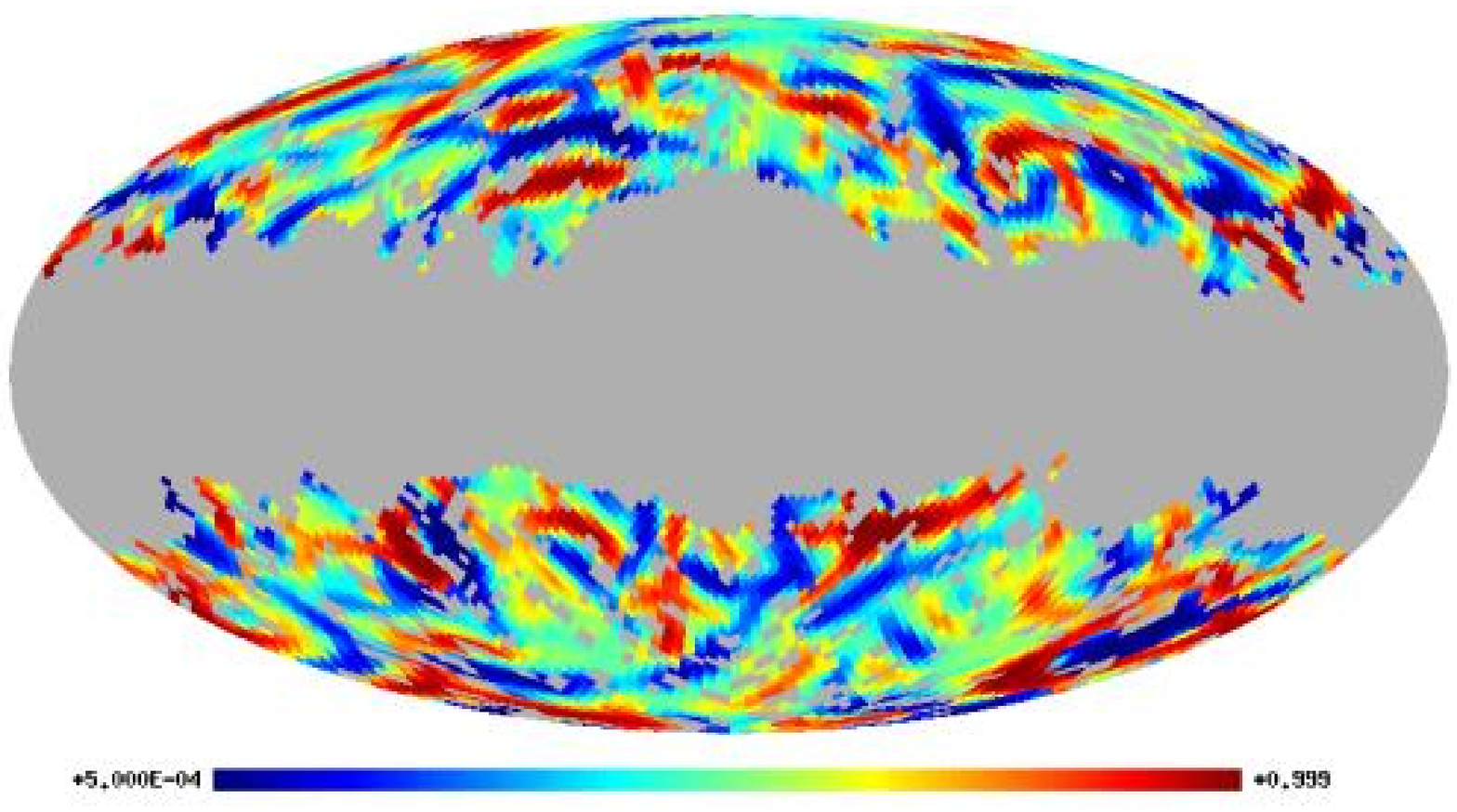}\includegraphics[width=4cm,keepaspectratio]{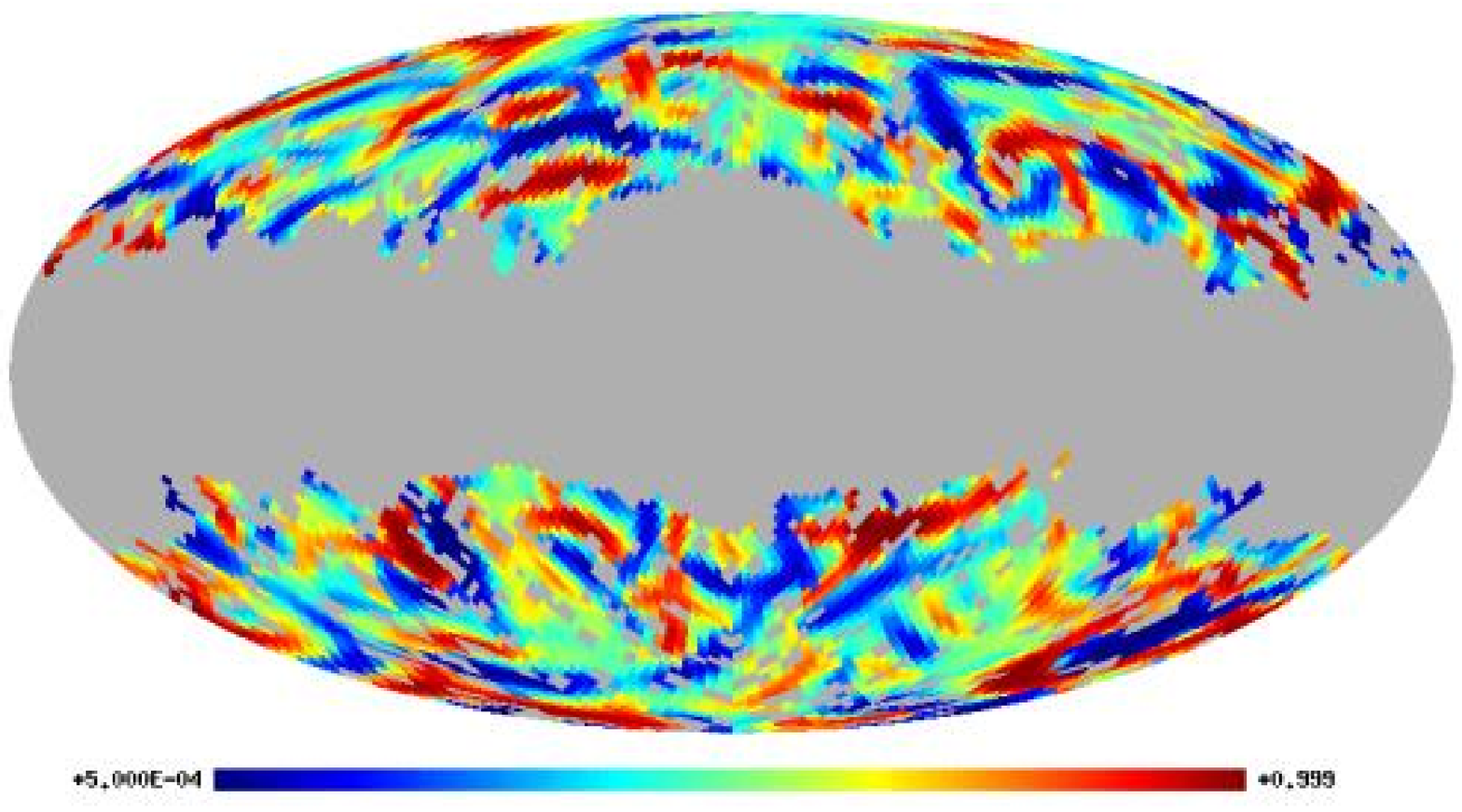}\includegraphics[width=4cm,keepaspectratio]{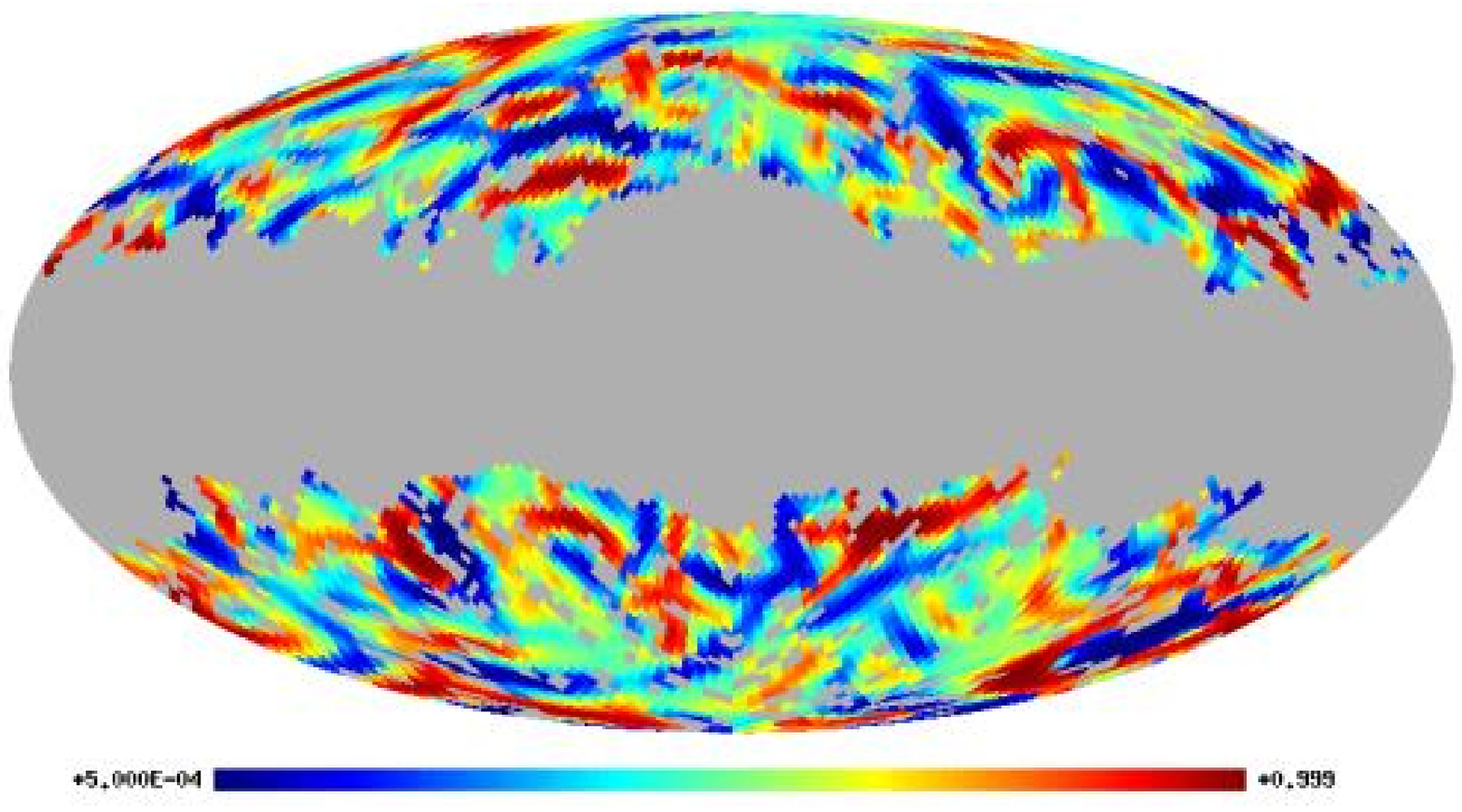}\includegraphics[width=4cm,keepaspectratio]{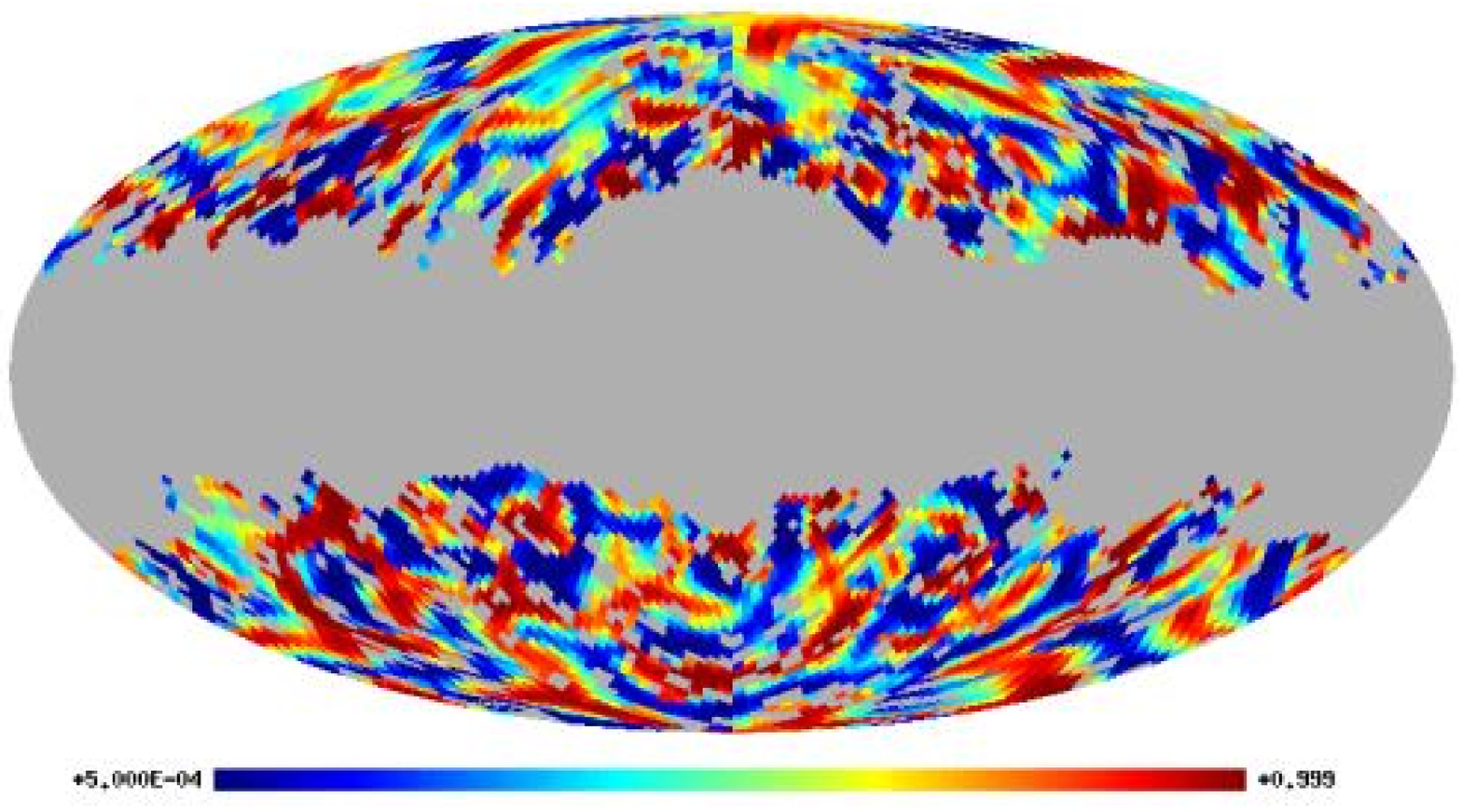}

\caption{\label{fig:fore}Alignment and signed-intensity analyses: cumulative
probabilities maps in Mollweide projection for the total weights $TW^{T}(\omega,a_{3})$
(top panel) and signed-intensities $I^{T}(\omega,a_{3})$ (bottom
panel) at wavelet scale $a_{3}$, resulting from the frequency maps
WCM-Q (extreme left column), WCM-V (center-left column), WCM-W (center-right
column), and the difference map WCM-nWVQ (extreme right column). The
total weights and signed-intensities patterns obtained from the WCM-Q,
WCM-V, and WCM-W maps are similar to those obtained from the WCM123
map. Foreground emissions can therefore be discarded as a possible
origin of the alignment and signed-intensity detections. The total
weights and signed-intensities patterns obtained from the WCM-nWVQ
map do not correspond to those obtained from the WCM123 map. This
is an additional evidence that instrumental noise and foreground emissions
can be discarded as possible origins of the alignment and signed-intensity
detections.}
\end{figure*}

A possible origin in terms of residual foreground emissions in the
WCM123 map is analyzed by reproducing the analyses on the frequency
maps WCM-Q, WCM-V, and WCM-W, which contain essentially CMB, with
instrumental noise and residual foreground emissions (see Subsection
\ref{sub:datasimul}). The analyses are also reproduced on the difference
map WCM-nWVQ which essentially contains instrumental noise and residual
foreground emissions (see Subsection \ref{sub:datasimul}). The corresponding
maps representing the cumulative probabilities that the simulations
exhibit a lower value than the data are presented in Figure \ref{fig:fore},
both for the total weights $TW^{T}(\omega,a_{3})$ and signed-intensities
$I^{T}(\omega,a_{3})$ at wavelet scale $a_{3}$. The total weights
and signed-intensities patterns obtained from the WCM-Q, WCM-V , and
WCM-W maps are similar to those obtained from the WCM123 map on which
the anomaly is observed. Consequently, the detection is independent
of the frequency. As foreground emissions themselves are frequency-dependent,
they may be discarded as a possible origin of the alignment and signed-intensity
detections reported above. Moreover, the total weights and signed-intensities
patterns obtained from the WCM-nWVQ maps do not correspond to those
obtained from the WCM123 map. This is an additional evidence that
instrumental noise and foreground emissions can be discarded as possible
origins of the alignment and signed-intensity detections.%

\subsection{Unknown systematics}

In the absence of instrumental noise or foreground emissions explanations,
the fact that the ecliptic poles axis and the CMB dipole axis are
strongly highlighted through our detections as preferred axes in the
sky still suggests the possible presence of unknown systematics, more
than a cosmological origin of the detections. These directions are
indeed local concepts with \emph{a priori} no global cosmological
meaning.

The wavelet analysis allowed us to identify the scales, roughly around
$10^{\circ}$ of angular size, at which the anomalies are observed.
This represents an important complementary piece of information in
the search for their origin. Let us recall in that respect that the
size of the mesh of the WMAP scanning pattern defined by the combination
of the spin and precession of the satellite is of the order of several
degrees \citep{bennett03c}. Moreover, the WMAP scanning strategy is
also obviously connected to the ecliptic coordinates system. Unknown
systematics related to the WMAP scanning strategy are therefore a
plausible source of the detected anomaly.

In that vein, it was recently suggested that North-South asymmetries
can in principle be explained by a bad calibration of the CMB dipole
\citep{freeman06}. More generally it was also suggested that a modulation
$T(\omega)\times[1+f(\omega)]$ of the CMB temperature field $T(\omega)$
with a modulation function $f(\omega)$, potentially coming from systematic
effects, could in turn explain North-South asymmetries and low multipoles
alignment \citep{helling06,gordon07}. A dipolar modulation function\emph{,
i.e.} containing only the multipole $l=1$, which provides the best
fit between the three-year WMAP data and the concordance model was
proposed \citep{eriksen07}. A best-fit dipolar-quadrupolar modulation
function\emph{,} \emph{i.e.} containing the multipoles $l=1$ and
$l=2$, was also proposed \cite[see arXiv:astro-ph/0603449v1]{spergel07}.
A possible origin of our anomalies in terms of a dipolar or dipolar-quadrupolar
modulation is probed by reproducing the alignment and signed-intensity
analyses on the WCM123 map correspondingly corrected%
\footnote{Let us remark that the modulations used were not primarily computed
for the WCM123 itself. The dipolar modulation considered was produced
by \citep{eriksen07} for the Internal Linear Combination (ILC) map
defined in \citep{hinshaw07}, at HEALPix resolution $N_{side}=16$.
The dipolar-quadrupolar modulation considered was produced by \cite[see arXiv:astro-ph/0603449v1]{spergel07}
for the frequency map WCM-V with Kp2 mask at HEALPix resolution $N_{side}=8$.
However, the best-fit modulations were shown not to be sensitive to
the three-year WMAP data set and sky cut. They are therefore also
adequate for correction of the WCM123 map.%
}.

Firstly, the results induced by the dipolar correction are as follows.
The alignment anomaly is preserved at the wavelet scales $a_{3}$
and $a_{4}$. The general pattern of anomalous directions still highlights
the same two mean preferred planes and the same mean preferred axis
in the sky. The global significance level of detection is not significantly
modified at neither of the two scales. The signed-intensity anomaly
is decreased at all the wavelet scales $a_{3}$, $a_{4}$, $a_{5}$,
and $a_{6}$, even though the general pattern of anomalous directions
still highlights the same three spots in the southern galactic hemisphere.
The axis of the dipolar modulation has its southern end at $(\theta,\varphi)=(117^{\circ},225^{\circ})$,
close to the cold spot centered at $(\theta,\varphi)=(150^{\circ},209^{\circ})$.
The correction consequently reduces the number of anomalous pixels
which define that spot, thereby reducing its impact on the anomaly
\citep{eriksen07}. The other two spots are essentially left unchanged
as they lie in the perpendicular plane, where the dipolar modulation
is negligible. The global significance level of detection is increased
at all scales. Typically, at the main wavelet scale of detection $a_{3}$,
the global significance level rises from $1.39\%$ to $4.39\%$.

Secondly, the results induced by the dipolar-quadrupolar correction
are as follows. The alignment anomaly is enhanced at the wavelet scales
$a_{3}$ and $a_{4}$. The general pattern of anomalous directions
still highlights the same two mean preferred planes and the same mean
preferred axis in the sky. The global significance level of detection
is significantly decreased at both scales. Typically, at the main
wavelet scale of detection $a_{3}$, the global significance level
drops from $0.85\%$ to $0.19\%$. The distribution of the signed-intensity
anomalies on the celestial sphere is affected at all the wavelet scales
$a_{3}$, $a_{4}$, $a_{5}$, and $a_{6}$, even though the general
pattern of anomalous directions still highlights the same three spots
in the southern galactic hemisphere. The correction pattern is such
that it still reduces the number of anomalous pixels which define
the cold spot centered at $(\theta,\varphi)=(150^{\circ},209^{\circ})$,
while enhancing the number of pixels defining the hot spot. The second
cold spot is essentially left unchanged. The global significance level
of detection is not significantly modified at none of the scales. 

Consequently, the modulations considered clearly fail to explain the
alignment and signed-intensity anomalies. Moreover the impact of the
correction is extremely sensitive to the arbitrary choice of the dipolar
or dipolar-quadrupolar nature of the modulation. The existence of
such a modulation, and its precise pattern, still need to be physically
verified in terms of specific systematic effects before full credit
may be given to the modifications induced.

In conclusion, neither instrumental noise nor foreground emissions
seem to explain the alignment and signed-intensity anomalies reported.
And first attempts to consider unknown systematics, potentially related
to the WMAP scanning strategy, also failed to explain our detections.
Consequently, a cosmological origin imprinted in the CMB can actually
not be discarded. In that regard, as already emphasized \citep{wiaux06a},
let us notice that the average wavelet scale at which our detections
are made, around ${10}^{\circ}$ of angular size on the celestial
sphere, is compatible with the size of primary CMB anisotropies due
to topological defects such as texture fields \citep{turok90} or secondary
anisotropies due to the Rees-Sciama effect \citep{martinez90}.

\section{Conclusion}

\label{sec:conclusion}

A significant statistical anisotropy was detected on the three-year
WMAP data of the CMB, through a decomposition of the signal with steerable
wavelets on the sphere.

Firstly, the alignment analysis of local CMB features performed on
the three-year WMAP data provides further insight on results obtained
from the one-year WMAP data analysis \citep{wiaux06a}. At a wavelet
scale corresponding to an angular size of ${8.3}^{\circ}$ on the
celestial sphere, a peculiar pattern of directions anomalous at $99.865\%$
($\pm3\sigma$ Gaussian) is detected, with a global significance level
of $0.85\%$. Two mean preferred planes are identified in the sky,
whose normal axes lie close to the CMB dipole axis. The first plane
is defined by the directions toward which the local CMB features are
anomalously aligned. In this plane, a prominent cluster of anomalous
directions defines a mean preferred axis very close to the ecliptic
poles axis. The second plane is defined by the directions anomalously
avoided by the local CMB features.

Secondly, the signed-intensity analysis of local CMB features performed
on the three-year WMAP data allows one to detect, at the same wavelet
scale, another specific pattern of directions anomalous at $99.865\%$,
with a global significance level of $1.39\%$. It points out three
clusters of anomalously high or low temperature in the southern galactic
hemisphere, thereby defining three mean preferred directions in the
sky. A first one essentially identifies with a known cold spot \citep{vielva04}.
A second one is a cold spot very close to the southern end of the
CMB dipole axis. The third one is a hot spot close to the southern
end of the ecliptic poles axis. 

Both detections are confirmed at neighbour wavelet scales corresponding
to slightly larger angular sizes on the celestial sphere. A robust
all-scale significance level is also estimated which accounts at once
for the twelve wavelet scales probed and for the detections observed
at various scales in both the alignment and signed-intensity analyses.
Its value is of $1.50\%$, which confirms the best levels of detections
observed in the data in terms of global significance levels at individual
wavelet scales in each of the two analyses.

Finally, we analyzed a possible relation between both anomalies, and
probed their possible origins. The hot spot of the signed-intensity
anomaly is partially, but certainly not totally, responsible for the
part of the alignment anomaly associated with the directions toward
which the local CMB features are anomalously aligned. Hence, the alignment
and signed-intensity anomalies observed are only very partially related.
Further analyses on frequency maps and differences of frequency maps
strongly reject instrumental noise and foreground emissions as possible
origins of our detections. Nevertheless, the fact that the detection
once more highlights the ecliptic poles axis and the CMB dipole axis
as preferred axes in the sky still calls for an explanation of the
origin of the anomalies in terms of systematics, possibly related
to the WMAP scanning strategy. We performed a simple test in that
context, certainly not exhaustive though, following recent suggestions
that unknown systematics might induce a modulation of the CMB temperature
field. But correcting the three-year WMAP data for the proposed dipolar
or dipolar-quadrupolar modulations fails to explain the detected anomalies.

A cosmological origin, that is, a global violation of the isotropy
of the Universe inducing an intrinsic statistical anisotropy of the
CMB, therefore still remains a plausible explanation of the detected
alignment and signed-intensity anomalies.

\section*{Acknowledgments}

The work of P.V. is funded through an I3P contract from the Spanish
National Research Council (CSIC). P.V. and E.M.-G. are also supported
by the Spanish MCYT project ESP2004-07067-C03-01. The work of Y.W.
is funded by the Swiss National Science Foundation (SNF) under contract
No. 200021-107478/1. Y.W. is also postdoctoral researcher of the Belgian
National Science Foundation (FNRS). The authors acknowledge the use
of the Legacy Archive for Microwave Background Data Analysis (LAMBDA).
Support for LAMBDA is provided by the NASA Office of Space Science.
The authors also acknowledge the use of the HEALPix and CAMB softwares,
and of the Univiewer visualization program.

\label{lastpage}

\end{document}